\documentclass[11pt]{article}
\usepackage[numbers]{natbib}
\usepackage{a4}
\usepackage{amsmath}
\usepackage{amssymb}
\usepackage{latexsym}
\usepackage{amssymb}
\usepackage{epsfig}
\usepackage{natbib}
\usepackage{bm}
\def \ii{{\mathrm{i}}}

\def \d{{\mathrm{d}}}

\def \pd{\partial}

\def \Bs{\boldsymbol{s}}

\def \BV{{\boldsymbol{V}}}

\def \BJ{{\boldsymbol{J}}}

\def \rr{{\boldsymbol{r}}}

\def \BB{{\boldsymbol{B}}}
\def \BE{{\boldsymbol{E}}}
\def \BH{{\boldsymbol{H}}}
\def \BD{{\boldsymbol{D}}}
\def \BA{{\boldsymbol{A}}}

\def \tR{{t_{\text{R}}}}

\def\onedot{$\mathsurround0pt\ldotp$}
\def\cdddot#1{
  \mathbin{\vcenter{\baselineskip.67ex
    \hbox{\onedot}\hbox{\onedot}\hbox{\onedot}%
  }}%
}

\begin{document}
\title{{\bf 
Second gradient electrodynamics: 
Green functions, wave propagation, regularization and self-force}}
\author{
Markus Lazar~$^\text{}$\footnote{
{\it E-mail address:} lazar@fkp.tu-darmstadt.de (M.~Lazar).
}
\\ \\
        Department of Physics,\\
        Darmstadt University of Technology,\\
        Hochschulstr. 6,\\      
        D-64289 Darmstadt, Germany
}

\date{}    
\maketitle

{\sf Wave Motion~{\bf 95} (2020), 102531; 
https://doi.org/10.1016/j.wavemoti.2020.102531}
\\

\begin{abstract}
In this work,  
the theory of second gradient electrodynamics, which is  an important example of generalized electrodynamics,
is proposed and investigated. 
Second gradient electrodynamics is a  gradient field theory  with up to 
second-order derivatives of the electromagnetic field
strengths in the Lagrangian density. 
Second gradient electrodynamics possesses a weak nonlocality in space and
time.  
In the framework of second gradient electrodynamics,
the retarded Green functions, first-order derivatives of the retarded Green functions,
retarded potentials, retarded electromagnetic field strengths, 
generalized Li{\'e}nard-Wiechert potentials and the corresponding electromagnetic field strengths 
are derived for three, two and one spatial  dimensions.
The behaviour of the electromagnetic fields is investigated
on the light cone.  
In particular, the retarded Green functions and their first-order
derivatives show oscillations around the classical solutions inside the forward light cone and 
it is shown that they are singularity-free and regular on the light cone in three, two and one spatial  dimensions. 
In second gradient electrodynamics, the self-force and the energy release rate are 
calculated and the equation of motion of a charged point particle, 
which is an integro-differential equation where the infamous third-order time-derivative of the position does not appear, 
is determined. 
\\ 

\end{abstract}


\section{Introduction}

The Maxwell theory of electrodynamics is a powerful field theory for the electromagnetic fields 
and the prototype for any physical field theory and gauge field theory~\citep{Jackson,HO}.
However, the Maxwell electrodynamics is a classical continuum field theory which is not valid at short distances. 
In particular, the classical Maxwell electrodynamics has some important drawbacks: the electromagnetic fields possess singularities,
the self-energy and self-force of a point charge become infinite, 
the infamous $4/3$-problem of the electromagnetic mass in the Abraham-Lorentz theory, 
and the runaway solutions of the classical Lorentz-Dirac equation of motion.

There are at least two ways to obtain singularity-free fields in electrodynamics. 
The first one is based on nonlinear electrodynamics as proposed by~\citet{Born1} and \citet{Born2}.
In the so-called Born-Infeld electrodynamics~\citep{Born2}, the inhomogeneous ``Maxwell"-like equations
are nonlinear partial differential equations.
The nonlinear Born-Infeld theory represents a classical generalization of the Maxwell theory for accommodating stable solutions
for the description of ``electrons". 
No standard methods for solving such nonlinear partial differential equations are known, 
the superposition principle for the electromagnetic fields no longer holds,
and the method of Green functions is not applicable. Only ``particular" solutions
for point charges are known in the Born-Infeld theory. 
The solutions of the electromagnetic fields for a non-uniformly moving point charge and the Li\'enard-Wiechert type potentials 
are unknown in the Born-Infeld electrodynamics due to its nonlinear character.
The other one is the theory of gradient electrodynamics as independently proposed by~\citet{Bopp} and \citet{Podolsky}.
The so-called Bopp-Podolsky electrodynamics is a linear field theory of first gradient electrodynamics 
including one characteristic length scale parameter, $\ell$, the so-called Bopp-Podolsky length parameter. 
The Bopp-Podolsky electrodynamics is a generalized electrodynamics with linear
field equations of fourth order for the electromagnetic potentials
and is free of classical divergences. 
Using the Bopp-Podolsky electrodynamics, 
it  was possible to solve the $4/3$-problem~\citep{Frenkel}, 
and  to eliminate runaway solutions from the Lorentz-Dirac equation of motion~\citep{Frenkel99}.
As argued by \citet{Iwan,Kvasnica} and \citet{Cuzi}, the Bopp-Podolsky 
length scale parameter $\ell$ is in the order of $\sim 10^{-15}$ m, which is the order of the classical electron radius.
An important aspect  of the Bopp-Podolsky electrodynamics 
is that it gives a regularization 
of the Maxwell electrodynamics based on higher-order partial differential equations.
On the other hand, \citet{GP} have shown that \citet{Podolsky} did not use a proper gauge fixing condition in his theory, 
since he used the classical Lorentz gauge condition, leading to spurious results, and that a generalized 
Lorentz gauge condition must be used  in the Bopp-Podolsky electrodynamics (see also \citep{Lazar19}).

Because the Bopp-Podolsky electrodynamics is a linear field theory, 
the partial differential equations of fourth order can be solved 
using the method of Green functions.
The retarded electromagnetic potentials were given by~\citet{Lande} (see also
\citep{Perlick2015,Lazar19}). 
In the Bopp-Podolsky electrodynamics, the  Li\'enard-Wiechert type potentials and corresponding electromagnetic fields have been
given by~\citet{Perlick2015} and \citet{Lazar19}. 
 The retarded Bopp-Podolsky Green function and its first-order
derivatives show decreasing oscillations inside the forward light cone.
The behaviour of the electromagnetic potentials and electromagnetic field
strengths on the light cone is obtained from the behaviour of the Green
function and its first-order derivatives in the neighbourhood of the light
cone. The one-dimensional electric field of the Bopp-Podolsky
electrodynamics is singularity-free on the light cone. 
The two-dimensional and the three-dimensional electromagnetic field strengths in the Bopp-Podolsky
electrodynamics possess weaker singularities than the classical singularities 
of the electromagnetic field strengths in the Maxwell electrodynamics. 
In order to regularize the two-dimensional and three-dimensional electromagnetic field strengths in 
the Bopp-Podolsky electrodynamics towards singular-free fields 
on the light cone, 
generalized electrodynamics of higher order might be used~\citep{Lazar19}.
On the other hand, some aspects of theories of gradient electrodynamics of higher order have been
discussed by~\citet{Pais, Kvasnica2} and \citet{Treder}.

The aim of the present work is to derive the theory of second gradient electrodynamics as 
straightforward generalization of the Bopp-Podolsky electrodynamics (first gradient electrodynamics). 
Therefore, the present work is a generalization of the results obtained in~\citep{Lazar19} towards the theory
of second gradient electrodynamics with new physical results and insights into generalized electrodynamics. 
Moreover, second gradient electrodynamics is the linear, local extension of the Maxwell electrodynamics with up to second-order derivatives of the 
electromagnetic field strengths which is both Lorentz and gauge invariant.   
The motivation for a second gradient electrodynamics is to obtain singularity-free electromagnetic fields at the light cone.
In particular, we study the radiation theory (generalized Li{\'e}nard-Wiechert potentials and the electromagnetic field strengths
of a non-uniformly moving charged particle) in the framework of second gradient electrodynamics. 
The main purpose of this work is to give the retarded Green functions, the wave propagation, the electromagnetic fields  and the self-force 
in the second gradient electrodynamics. 
In Section~\ref{sec2}, we give the basics of second gradient electrodynamics.
In Section~\ref{sec3}, we give the collection of the retarded Green
functions and their first-order derivatives in three, two and one spatial  dimensions (3D, 2D, 1D).
The retarded potentials and retarded electromagnetic field strengths are given
in Section~\ref{sec4} for 3D, 2D and 1D. 
In Section~\ref{sec5}, the generalized Li{\'e}nard-Wiechert potentials and electromagnetic field strengths
in generalized Li{\'e}nard-Wiechert form are presented.
In Section~\ref{sec6}, the self-force, the energy release rate and the
equation of motion of a charged particle are given.
The conclusion are presented in Section~\ref{sec7}.

\section{Theory of second gradient electrodynamics}
\label{sec2}

In this section, we formulate the basic framework of the theory of second gradient electrodynamics
which is an important example of  generalized electrodynamics.

In the theory of second gradient electrodynamics, 
the Lagrangian density depends in addition to the classical Maxwell term 
also on both first- and second-order derivatives of the electromagnetic field strengths.
Therefore, the Lagrangian density of second gradient electrodynamics has the form 
\begin{align}
\label{L-BP}
{\cal L_{\text{grad}}}&=
\frac{\varepsilon_0}{2} 
\Big(\bm E \cdot \bm E 
+\ell_1^2 
\Big[
\nabla \bm E :\nabla \bm E 
-\frac{1}{c^2}\, \pd_t \bm E \cdot \pd_t \bm E 
\Big]\nonumber\\
&\quad
+\ell_2^4 
\Big[
\nabla\nabla \bm E \mathbin{\vdots} \nabla\nabla \bm E
-\frac{2}{c^2}\, \pd_{t} \nabla \bm E : \pd_{t} \nabla \bm E 
+\frac{1}{c^4}\, \pd_{tt} \bm E \cdot \pd_{tt} \bm E 
\Big]
\Big)
\nonumber\\
&\quad
-\frac{1}{2\mu_0 } 
\Big(\bm B \cdot \bm B 
+\ell_1^2 
\Big[
\nabla \bm B :\nabla \bm B 
-\frac{1}{c^2}\, \pd_t \bm B \cdot \pd_t \bm B 
\Big]\nonumber\\
&\quad
+\ell_2^4 
\Big[
\nabla\nabla \bm B \mathbin{\vdots} \nabla\nabla \bm B
-\frac{2}{c^2}\, \pd_{t} \nabla \bm B : \pd_{t} \nabla \bm B 
+\frac{1}{c^4}\, \pd_{tt} \bm B \cdot \pd_{tt} \bm B 
\Big]
\Big)
-\rho\phi+\bm J\cdot \bm A\,,
\end{align}
where the following notation has been used 
$ \nabla\nabla \bm E \mathbin{\vdots} \nabla\nabla \bm E 
=\pd_k\pd_j E_i \pd_k \pd_j E_i$,
$ \nabla \bm E :\nabla \bm E =\pd_j E_i \pd_j E_i$ and
$ \bm E \cdot\bm E =E_i E_i$. 
Here 
$\phi$ and $\BA$ are the electromagnetic gauge potentials,
$\BE$ is the electric field strength vector,
$\BB$ is the magnetic field strength vector,
$\rho$ is the electric charge density,
$\BJ$ is the electric current density vector,
$\varepsilon_0$ is the electric constant (or permittivity of vacuum)
and 
$\mu_0$ is the magnetic constant 
(or permeability of vacuum).
The speed of light in vacuum is defined by
\begin{align}
c=\frac{1}{\sqrt{\varepsilon_0\mu_0}}\,.
\end{align}
Moreover, $\ell_1$ and $\ell_2$ are 
the two (positive and real) characteristic length scale parameters in second gradient 
electrodynamics,
$\pd_t$ is the differentiation with respect to the time $t$ and
$\nabla$ is the Nabla operator.
In addition to the classical terms, first- and second-order 
spatial- and time-derivatives of the electromagnetic field strengths   ($\BE$, $\BB$) 
multiplied by the characteristic lengths $\ell_1$ and $\ell_2$ appear
in Eq.~\eqref{L-BP} which describe a weak nonlocality in space and time.
In fact, $\ell_1$ is the length parameter corresponding to first-order derivatives of the electromagnetic field strengths, 
while $\ell_2$ is the length parameter corresponding to second-order derivatives of the electromagnetic field strengths.
The limit $\ell_2^4\rightarrow 0$ in Eq.~\eqref{L-BP} provides the limit
of second gradient electrodynamics to the Bopp-Podolsky electrodynamics.

As usual, the electromagnetic field strengths ($\BE$, $\BB$)
can be given in terms
of the electromagnetic gauge potentials (scalar potential $\phi$, vector potential $\BA$) according to 
\begin{align} 
\label{E}
\BE&=-\nabla \phi-\pd_t \bm A\,,\\
\label{B}
\BB&=\nabla\times \bm A\,.
\end{align}
Due to the definition of the electromagnetic field strengths~\eqref{E} and \eqref{B},
the two electromagnetic Bianchi identities are satisfied
\begin{alignat}{2}
\label{BI-1}
\nabla\times\BE+\pd_t\BB&=0\qquad\qquad
&&(\text{Faraday law})\,,\\
\label{BI-2}
\nabla\cdot \BB&=0\qquad\qquad
&&(\text{magnetic field closed})\,,
\end{alignat}
which are known as homogeneous Maxwell equations.

The Euler-Lagrange equations obtained from the Lagrangian~\eqref{L-BP} due to the variation with respect to the scalar potential $\phi$ and the 
vector potential $\bm A$, 
$\frac{\delta {\cal L_{\text{grad}}}}{\delta \phi}=0$
and 
$\frac{\delta {\cal L_{\text{grad}}}}{\delta \bm A}=0$,
give the electromagnetic field equations
\begin{align}
\label{EL-1}
& L(\square)\,
\nabla\cdot \bm E=\frac{1}{\varepsilon_0}\,\rho\,,\\
\label{EL-2}
&
 L(\square)
\Big(
\nabla\times\bm B-\frac{1}{c^2}\,\pd_t\bm E\Big)=\mu_0\,\BJ\,,
\end{align}
respectively.
The appearing differential operator of fourth order is given by
\begin{align}
\label{L-op}
 L(\square)=1+\ell_1^2 \square +\ell_2^4 \square^2\,.
\end{align}
The d'Alembert operator (or wave operator) is defined as 
\begin{align}
\square:=\frac{1}{c^2}\,\pd_{tt}-\Delta\,,
\end{align}
where $\Delta$ is the Laplace operator.
Eqs.~\eqref{EL-1} and \eqref{EL-2} represent the generalized inhomogeneous
Maxwell equations in second gradient electrodynamics.  
Of course, the electric current density vector 
and the electric charge density
satisfy the equation of continuity
\begin{align}
\label{CE}
&\nabla\cdot \BJ+\pd_t\rho=0\,.
\end{align}

Using the variational derivative with respect to 
the electromagnetic fields ($\bm E$, $\bm B$),
we obtain the spacetime relations (or constitutive equations for a vacuum) in second gradient electrodynamics 
for the response quantities
($\bm D$, $\bm H$) 
\begin{align}
\label{CE1}
\BD&:=\frac{\delta{\cal L_{\text{grad}}}}{\delta \BE}=\varepsilon_0\, 
 L(\square)\,
\BE\,,\\
\label{CE2}
\BH&:=-\frac{\delta{\cal L_{\text{grad}}}}{\delta \BB}=\frac{1}{\mu_0}\,
 L(\square)\,
\BB\,,
\end{align}
where $\BD$ is the electric excitation vector and 
$\BH$ is the magnetic excitation vector.
The higher order terms in Eqs.~\eqref{CE1} and \eqref{CE2} describe the polarization
of the vacuum present in second gradient electrodynamics. 
The Euler-Lagrange equations~\eqref{EL-1} and \eqref{EL-2} 
can be rewritten in the form of inhomogeneous Maxwell equations 
using the constitutive equations~\eqref{CE1} and \eqref{CE2}
\begin{alignat}{2}
\label{ME-inh}
\nabla\cdot \BD&=\rho\qquad\qquad
&&(\text{Gauss law})
\,,\\
\label{ME-inh2}
 \nabla\times\BH-\pd_t\BD&=\BJ\qquad\qquad
&&(\text{Oersted-Amp{\`e}re law})
\,. 
\end{alignat}

Moreover,
from Eqs.~(\ref{EL-1}) and (\ref{EL-2}) and using Eqs.~\eqref{BI-1} and \eqref{BI-2},
inhomogeneous partial differential equations, being partial differential
equations of sixth order, can be obtained for the electromagnetic field strengths
\begin{align}
\label{E-w}
 L(\square)\,
\square\,\BE&=-\frac{1}{\varepsilon_0}\Big(\nabla\rho+\frac{1}{c^2}\, \pd_t\BJ\Big)\,,\\
\label{B-w}
 L(\square)\,
\square\,\BB&=\mu_0\,\nabla\times\BJ\,.
\end{align}
If we take into account the generalized Lorentz gauge condition~\citep{GP,Lazar19}
\begin{align}
\label{LG}
 L(\square)\bigg(
\frac{1}{c^2}\,\pd_t \phi+ \nabla \cdot \bm A\bigg)=0\,,
\end{align}
the following 
inhomogeneous partial differential equations of sixth order
are obtained for the electromagnetic gauge potentials 
from Eqs.~(\ref{EL-1}) and (\ref{EL-2})
\begin{align}
\label{phi-w}
 L(\square)\,
\square\,\phi&=\frac{1}{\varepsilon_0}\, \rho\,,\\
\label{A-w}
L(\square)\,\square\,\bm A&=\mu_0\, \BJ\,.
\end{align}

The differential operator of fourth order~\eqref{L-op}  
can be written in the form as  product of two Klein-Gordon operators with 
two length scale parameters $a_1$ and $a_2$,
which is called bi-Klein-Gordon operator,
\begin{align}
\label{L-op-2}
L(\square)=\big(1+a_1^2\square\big)\big(1+a_2^2\square\big)
\end{align}
with
\begin{align}
\label{a1a2-1}
\ell_1^{2}&=a_1^{2}+a_2^{2}\, ,\\
\label{a1a2-2}
\ell_2^{4}&=a_1^{2}\, a_2^{2}\,
\end{align}
and 
\begin{align}
\label{a1-2}
a^{2}_{1,2}&=\frac{\ell_1^{2}}{2}\Bigg(1\pm\sqrt{1-4\,\frac{\ell_2^{4}}{\ell_1^{4}}}\Bigg)\,.
\end{align}

Using the two length scale parameters $a_1$ and $a_2$, two subsidiary masses
corresponding to the two Klein-Gordon operators can be introduced as
\begin{align}
\label{mass}
m_1=\frac{\hbar}{c a_1}\,,\qquad
m_2=\frac{\hbar}{c a_2}\,,
\end{align}
where  $\hbar$ is the reduced Planck constant.

In general, the length scale parameters $a_1$ and $a_2$, appearing in the
Klein-Gordon operators 
$(1+a_1^2\square)$ and $(1+a_2^2\square)$, 
may be real or complex.
In the theory of second gradient electrodynamics,  
the condition for the character, real or complex, of the lengths 
$a_1$ and $a_2$ can be obtained from the condition 
if the discriminant, $1-4\ell_2^4/\ell_1^4$, 
is positive or negative in Eq.~\eqref{a1-2}.
Depending on the character of the length scales $a_1$ and $a_2$, 
it can be  distinguished between the following cases:
\begin{itemize}
\item[(1)]  
$\ell_1^4>4\ell_2^4$\,:\\
In this case, $a_1$ and $a_2$ are real and distinct and they read as 
\begin{align}
\label{a1-2-2}
a_{1,2}&=\ell_1\,\sqrt{\frac{1}{2}\pm \frac{1}{2}\,\sqrt{1-4\left(\frac{\ell_2}{\ell_1}\right)^{\!4}}}
\end{align}
with $a_1>a_2$.
The limit to the Bopp-Podolsky electrodynamics  is: 
$\ell_2^4\rightarrow 0$, and therefore $a_1\rightarrow \ell$ and $a_2\rightarrow 0$. 
\item[(2)]
$\ell_1^4=4\ell_2^4$\,: \\
The lengths $a_1$ and $a_2$ are real and equal
\begin{align}
 a_1 =a_2=\frac{\ell_1}{\sqrt{2}}=\ell_2\,.
 \end{align}  
There is no limit to the Bopp-Podolsky electrodynamics.
This case can lead to Green functions having a time dependence that increases or decreases slowly, which can give rise to unphysical results
(see Section~\ref{GF-mul}).
\item[(3)]
$\ell_1^4<4\ell_2^4$\,:\\
The two lengths $a_1$ and $a_2$  are complex conjugate
\begin{align}
\label{c1-2-c}
a_{1,2}&=A\pm\ii B\,,
\end{align}
 with
\begin{align}
\label{AB}
A=\ell_2\,\sqrt{\frac{1}{2}+\frac{\ell_1^2}{4\ell_2^2}}\,,\qquad
B=\ell_2\,\sqrt{\frac{1}{2}-\frac{\ell_1^2}{4\ell_2^2}}\,.
\end{align}
There is no limit to the Bopp-Podolsky electrodynamics.
Furthermore, this case leads to Green functions having a time dependence that increases exponentially and leading to
acausal propagation  and to complex masses (e.g.~\citep{Pais,MS}). 
\end{itemize}

Therefore, the case~(1) is the physical one and is the generalization of the Bopp-Podolsky electrodynamics (first gradient electrodynamics)
towards second gradient electrodynamics.

\section{Green functions in second gradient electrodynamics}
\label{sec3}

In this section, we derive the retarded Green functions of 
second gradient electrodynamics. 
Second gradient electrodynamics is a linear field theory 
with partial differential equations of sixth order.
Therefore, the method of Green functions, which are fundamental solutions of linear partial differential operators,
can be used to derive exact analytical solutions.

The Green function $G^{L\square}$ of the bi-Klein-Gordon-d'Alembert operator,
being a sixth-order differential operator, 
$L(\square)\, \square$, is defined by 
\begin{align}
\label{BPE}
 L(\square)\,
\square\, G^{L\square}(\bm R, \tau)=\delta(\tau)\delta(\bm R)\,,
\end{align}
where $\tau=t-t'$, $\bm R= \bm r-\bm r'$ and $\delta$ is the Dirac
delta-function. 
Thus, the Green function,  $G^{L\square}$, 
is the fundamental solution of the linear hyperbolic differential operator 
of sixth order, $L(\square)\,\square$, in the sense of the distribution theory~\citep{Schwartz}.
For the retarded Green function, the causality constraint must be fulfilled
\begin{align}
\label{CC}
G^{L\square}(\bm R, \tau)=0\qquad \text{for}\quad \tau<0\,.
\end{align}
As always for hyperbolic operators, 
this is the only fundamental solution with support in the half-space $\tau\ge 0$ 
(see, e.g., \citep{Hoermander}).

On the other hand,
the partial differential equation of sixth order~(\ref{BPE}) 
might be written as an equivalent  system of partial differential equations of
lower order 
\begin{align}
\label{BPE-2}
 L(\square)\,
G^{L\square}(\bm R, \tau)&=G^\square(\bm R, \tau)\, ,\\
\label{BPE-3}
\square\, G^{L\square}(\bm R, \tau)&=G^{L}(\bm R, \tau)\, ,\\
\label{wave}
\square\, G^\square(\bm R, \tau)&=\delta(\tau)\delta(\bm R) \,,\\
\label{KGE}
 L(\square)\,
G^{L}(\bm R, \tau)&=\delta(\tau)\delta(\bm R)\,,
\end{align}
where $G^\square$ is the Green function of the d'Alembert operator, $\square$,
in Eq.~\eqref{wave} and 
$G^{L}$ is the Green function of the bi-Klein-Gordon operator, $L(\square)$,
in Eq.~\eqref{KGE}.
Moreover, it can be seen that the equation~(\ref{BPE}) is a bi-Klein-Gordon-d'Alembert equation.

If we use the partial fraction decomposition, 
then the inverse differential operators 
 $\big[L(\square)\big]^{-1}$ and
 $\big[L(\square) \square\big]^{-1}$  with Eq.~\eqref{L-op-2}
 read in the operator notation (see also \citep{Schwartz,Jager})
\begin{align}
\label{Deco-L}
\big[L(\square)\big]^{-1}=\frac{1}{a_1^2-a_2^2}
\Big(a_1^2\, \big[1+a_1^2 \square\big]^{-1}
-a_2^2\, \big[1+a_2^2 \square\big]^{-1}\Big)
\end{align}
and
\begin{align}
\label{Deco-L2}
\big[L(\square) \square\big]^{-1}=\square^{-1}
-\frac{1}{a_1^2-a_2^2}
\Big(a_1^4\, \big[1+a_1^2 \square\big]^{-1}
-a_2^4\, \big[1+a_2^2 \square\big]^{-1}\Big)\,.
\end{align}
Therefore,
the Green function $G^{L}$ might be
written as a linear combination of two Klein-Gordon Green functions $G^{\rm KG}(a_1)$ and  $G^{\rm KG}(a_2)$
corresponding to the two length scale parameters $a_1$ and $a_2$ and Klein-Gordon operators
$[1+a_1^2\square]$ and $[1+a_2^2\square]$
\begin{align}
\label{KGG}
G^{L}=\frac{1}{a_1^2-a_2^2}
\Big(a_1^2\, G^{{\rm KG}}(a_1)-a_2^2\, G^{\rm KG}(a_2)\Big)\,.
\end{align}
On the other hand, the Green function $G^{L\square}$ might be written 
as a linear combination of the Green function $G^\square$ of the
d'Alembert operator 
and  two Klein-Gordon Green functions $G^{\rm KG}(a_1)$ and  $G^{\rm KG}(a_2)$
corresponding to the two length scale parameters $a_1$ and $a_2$ 
\begin{align}
\label{BPG}
G^{L\square}=G^\square
-\frac{1}{a_1^2-a_2^2}
\Big(a_1^4\, G^{{\rm KG}}(a_1)-a_2^4\, G^{\rm KG}(a_2)\Big)\,.
\end{align}
Using Eq.~\eqref{KGG}, the Green function of the bi-Klein-Gordon equation 
can be derived from the expressions of 
the Green function of the Klein-Gordon equation  (see, e.g., \citep{Iwan,Zauderer,Poly}).
For that reason, 
the bi-Klein-Gordon field is a superposition of two Klein-Gordon
fields with the length scale parameters $a_1$ and $a_2$. 
Furthermore,
the Green function of the bi-Klein-Gordon-d'Alembert equation is derived by using the expressions of 
the Green function of the d'Alembert equation (see, e.g., \citep{Zauderer,Barton,Kanwal,Wl}) and
the Green function of the Klein-Gordon equation  (see, e.g.,
\citep{Iwan,Zauderer,Poly}) 
using Eq.~\eqref{BPG}.
Therefore, the  bi-Klein-Gordon-d'Alembert field is a superposition of the Maxwell field
and two Klein-Gordon fields. 

Moreover, the Green function of the bi-Klein-Gordon-d'Alembert equation may be
written as convolution of the Green function of the d'Alembert equation with 
the Green function of the bi-Klein-Gordon equation
\begin{align}
\label{GBP-conv}
G^{L\square}=G^\square*G^{L}\,,
\end{align}
satisfying Eqs.~\eqref{BPE}, \eqref{BPE-2} and \eqref{BPE-3}.
The symbol $*$ denotes the convolution in space and time.
Furthermore, the Green function of the bi-Klein-Gordon equation can be written as
convolution of the Green functions of the two Klein-Gordon equations (see also \citep{Jager,Kanwal})
\begin{align}
\label{GKG-conv}
G^{L}=
G^{{\rm KG}}(a_1)*G^{\rm KG}(a_2)\,,
\end{align}
satisfying Eqs.~\eqref{KGE} and \eqref{L-op-2}.
In Eq.~\eqref{GBP-conv}, 
the Green function $G^{L}$ 
of the bi-Klein-Gordon operator plays the role of the 
regularization function in second gradient electrodynamics, 
regularizing the  Green function $G^\square$ 
of the d'Alembert operator towards the Green function $G^{L\square}$.
On the other hand, the limit of $G^{L\square}$ as $a_1$ and $a_2$ tend to zero 
reads (see Eq.~\eqref{BPG})
\begin{align}
\label{lim-BPG}
\lim_{a_1 \to 0}\,\lim_{a_2 \to 0}
G^{L\square}=
\lim_{a_1 \to 0} G^\text{BP}
=G^\square\,
\end{align}
with 
\begin{align}
\label{lim-BPG-2}
\lim_{a_2 \to 0}
G^{L\square}=
G^\text{BP}\,,
\end{align}
where $G^\text{BP}$ is the Bopp-Podolsky Green function.

\subsection{3D Green functions}

The three-dimensional retarded Green functions of the d'Alembert operator~\eqref{wave}, the Klein-Gordon operator with length parameter $a_1$, 
the bi-Klein-Gordon operator~\eqref{KGE} and the bi-Klein-Gordon-d'Alembert
operator are given by
\begin{align}
\label{G-w-3d}
G_{(3)}^\square(\bm R, \tau)&=\frac{1}{4\pi R}\,\delta\big(\tau-R/c\big)\,,\\
\label{G-KG-3d}
G_{(3)}^{\rm KG}(\bm R, \tau)
&=\frac{1}{4\pi a_1^2}\,\bigg[
\frac{1}{R}\,\delta\big(\tau-R/c\big)
-\frac{c}{a_1}\,
\frac{H\big(c\tau-R\big)}{\sqrt{c^2 \tau^2-R^2}}\, 
J_1 \bigg( \frac{\sqrt{c^2 \tau^2-R^2}}{a_1}\bigg)\bigg]\,,\\
\label{G-BKG-3d}
G_{(3)}^{L}(\bm R, \tau)
&=-\frac{c}{4\pi(a_1^2-a_2^2)}\,
\frac{H\big(c\tau-R\big)}{\sqrt{c^2 \tau^2-R^2}}\, 
\bigg[
\frac{1}{a_1}\,J_1 \bigg( \frac{\sqrt{c^2 \tau^2-R^2}}{a_1}\bigg)
-\frac{1}{a_2}\,J_1 \bigg( \frac{\sqrt{c^2 \tau^2-R^2}}{a_2}\bigg)
\bigg]\,,\\
\label{G-BP-3d}
G_{(3)}^{L\square}(\bm R, \tau)
&=\frac{c}{4\pi(a_1^2-a_2^2)}\,
\frac{H\big(c\tau-R\big)}{\sqrt{c^2 \tau^2-R^2}}\, 
\bigg[
{a}_1 J_1 \bigg( \frac{\sqrt{c^2 \tau^2-R^2}}{a_1}\bigg)
-a_2 J_1 \bigg( \frac{\sqrt{c^2 \tau^2-R^2}}{a_2}\bigg)
\bigg]\,,
\end{align}
where $R=\sqrt{(x-x')^2+(y-y')^2+(z-z')^2}$, $H$ is the Heaviside step function 
and $J_1$ is the Bessel function of the first kind of order 1. 
Eq.~\eqref{G-BKG-3d} is obtained from Eq.~\eqref{KGG} using Eq.~\eqref{G-KG-3d}
and is non-singular, since the $\delta$-term, present in the Green function of the Klein-Gordon operator,
vanishes due to the superposition of the two Klein-Gordon Green functions.
The Green function~\eqref{G-BKG-3d} is in agreement with the expression given 
in~\citep{Rzewuski2,Jager,Kanwal}. 
On the other hand, Eq.~\eqref{G-BP-3d} is obtained from Eq.~\eqref{BPG} using the Green functions~\eqref{G-w-3d} and \eqref{G-KG-3d}
and is also non-singular.

\begin{figure}[tp]\unitlength1cm
\vspace*{-0.5cm}
\centerline{
\epsfig{figure=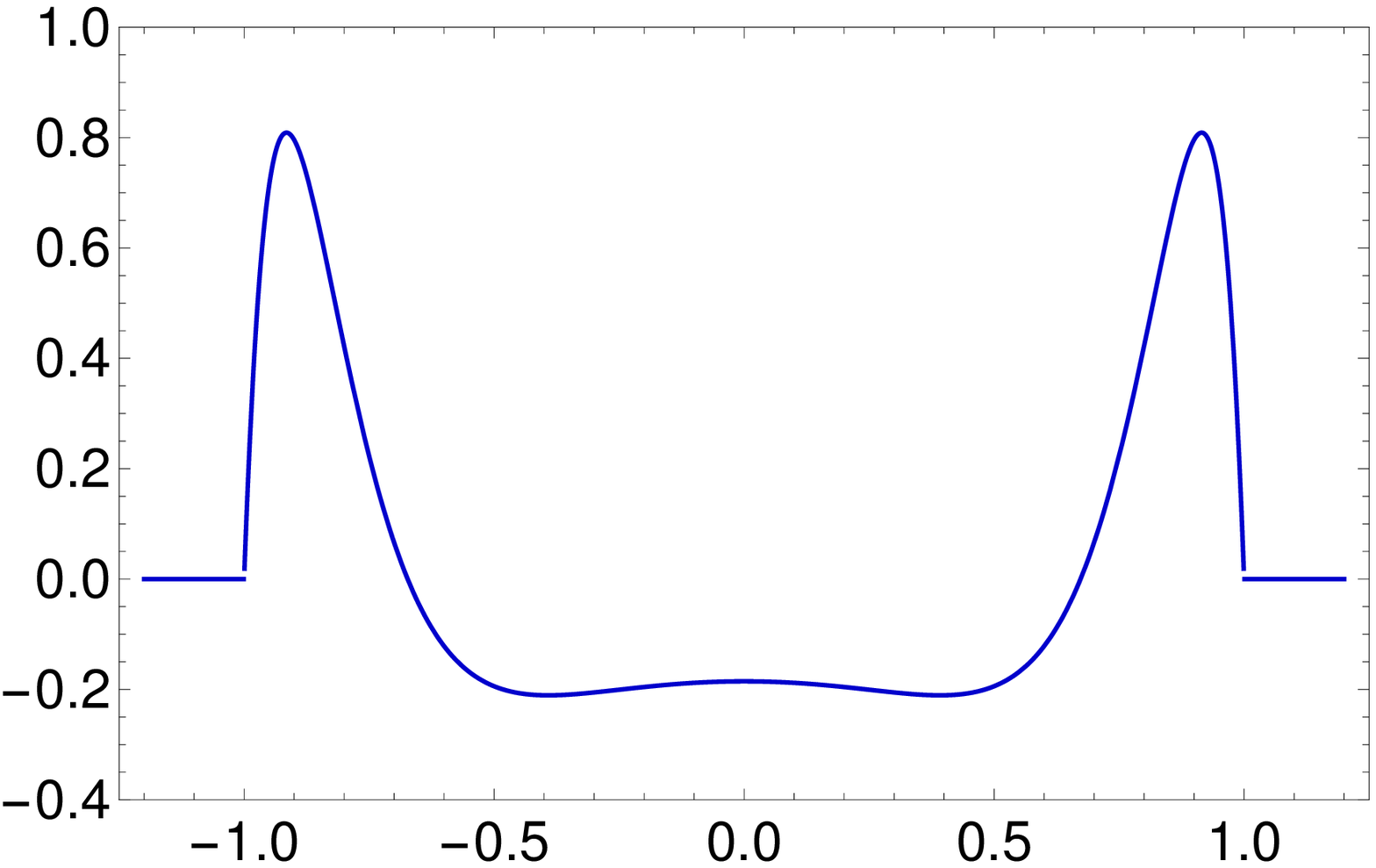,width=7.0cm}
\put(-3.5,-0.4){$X$}
\put(-7.1,-0.3){$\text{(a)}$}
\hspace*{0.4cm}
\put(-0.1,-0.3){$\text{(b)}$}
\put(3.5,-0.4){$\tau$}
\epsfig{figure=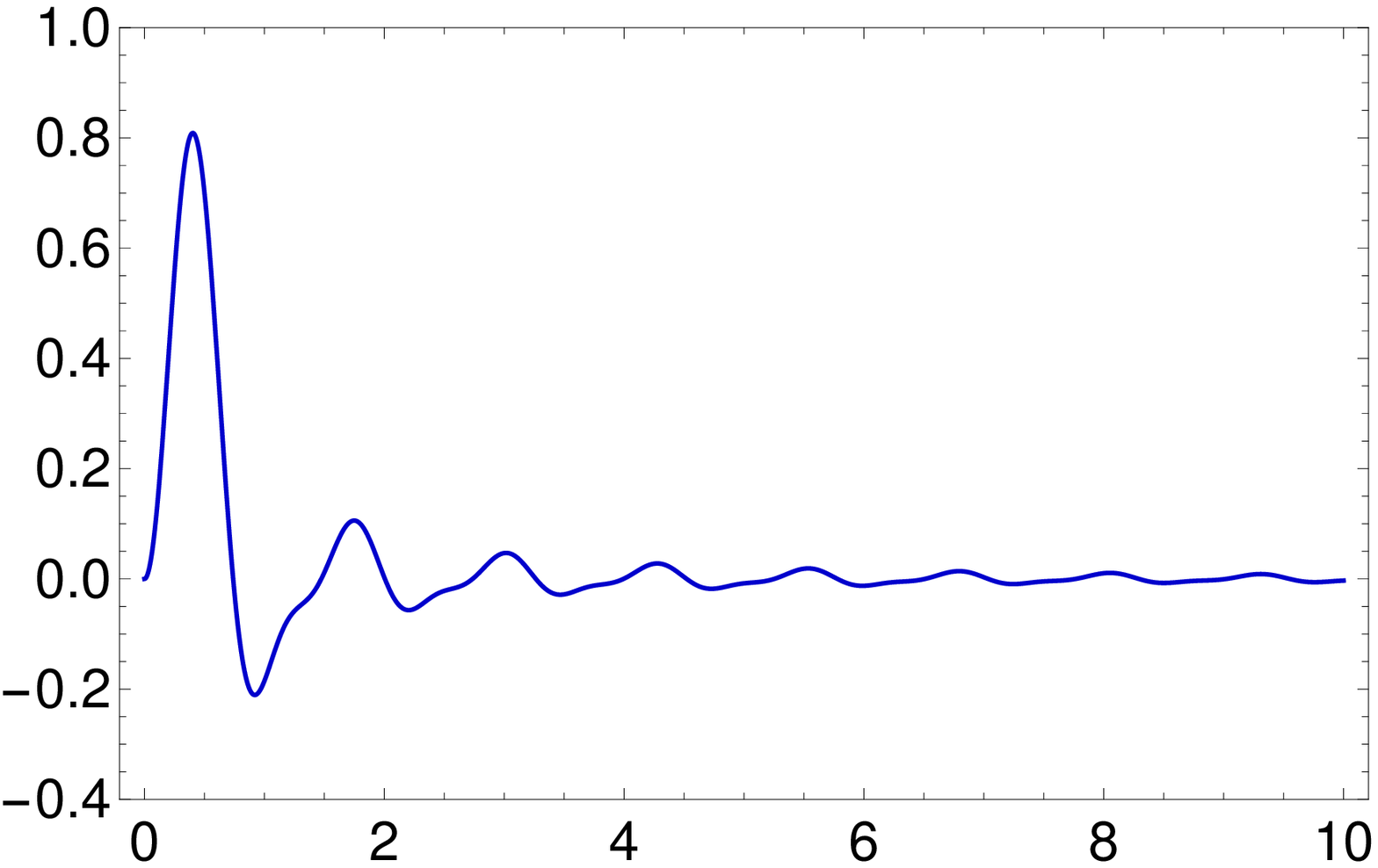,width=7.0cm}
}
\vspace*{0.2cm}
\centerline{
\epsfig{figure=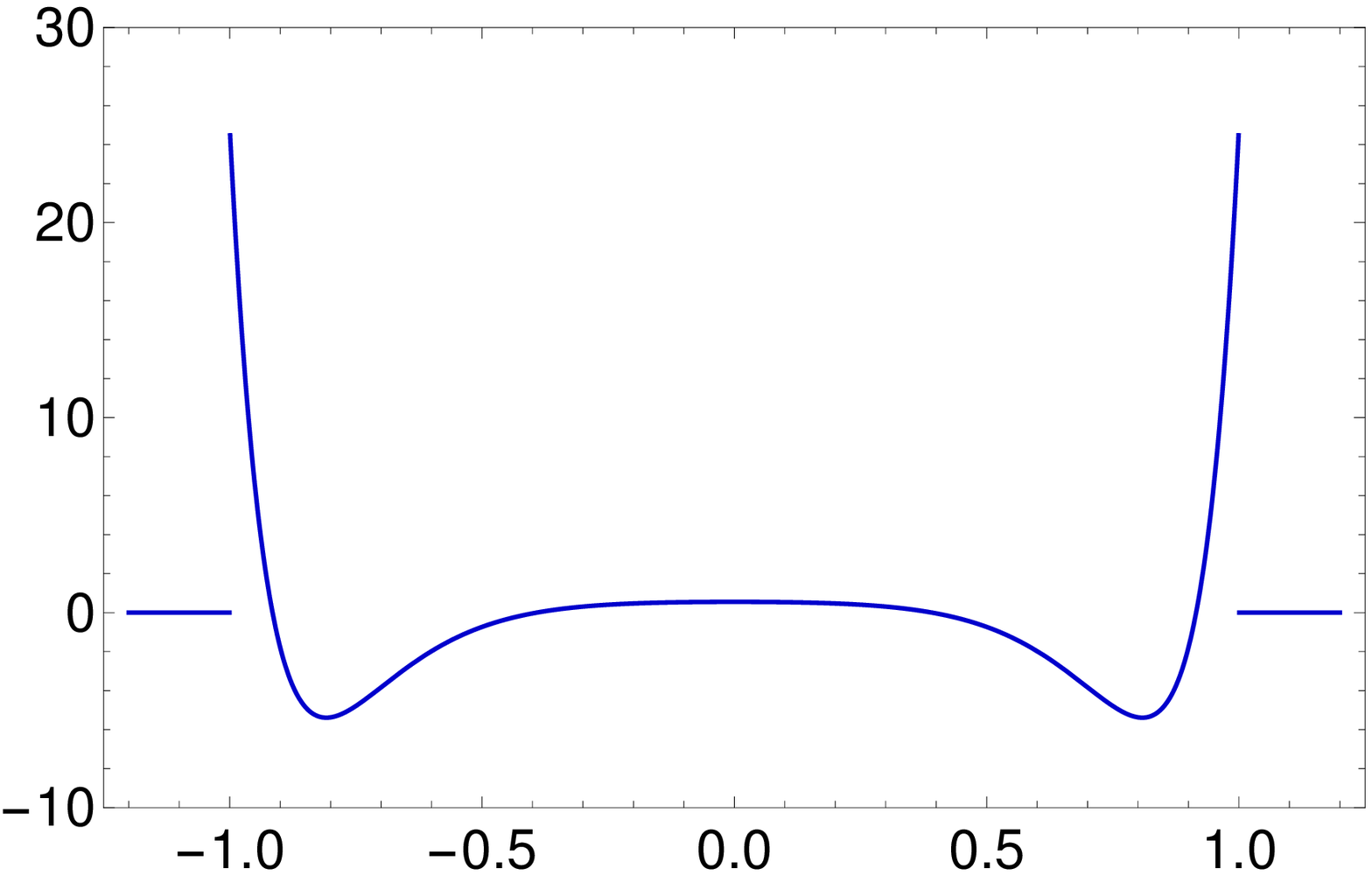,width=6.8cm}
\put(-3.5,-0.4){$X$}
\put(-7.1,-0.3){$\text{(c)}$}
\hspace*{0.4cm}
\put(3.5,-0.4){$\tau$}
\put(-0.1,-0.3){$\text{(d)}$}
\epsfig{figure=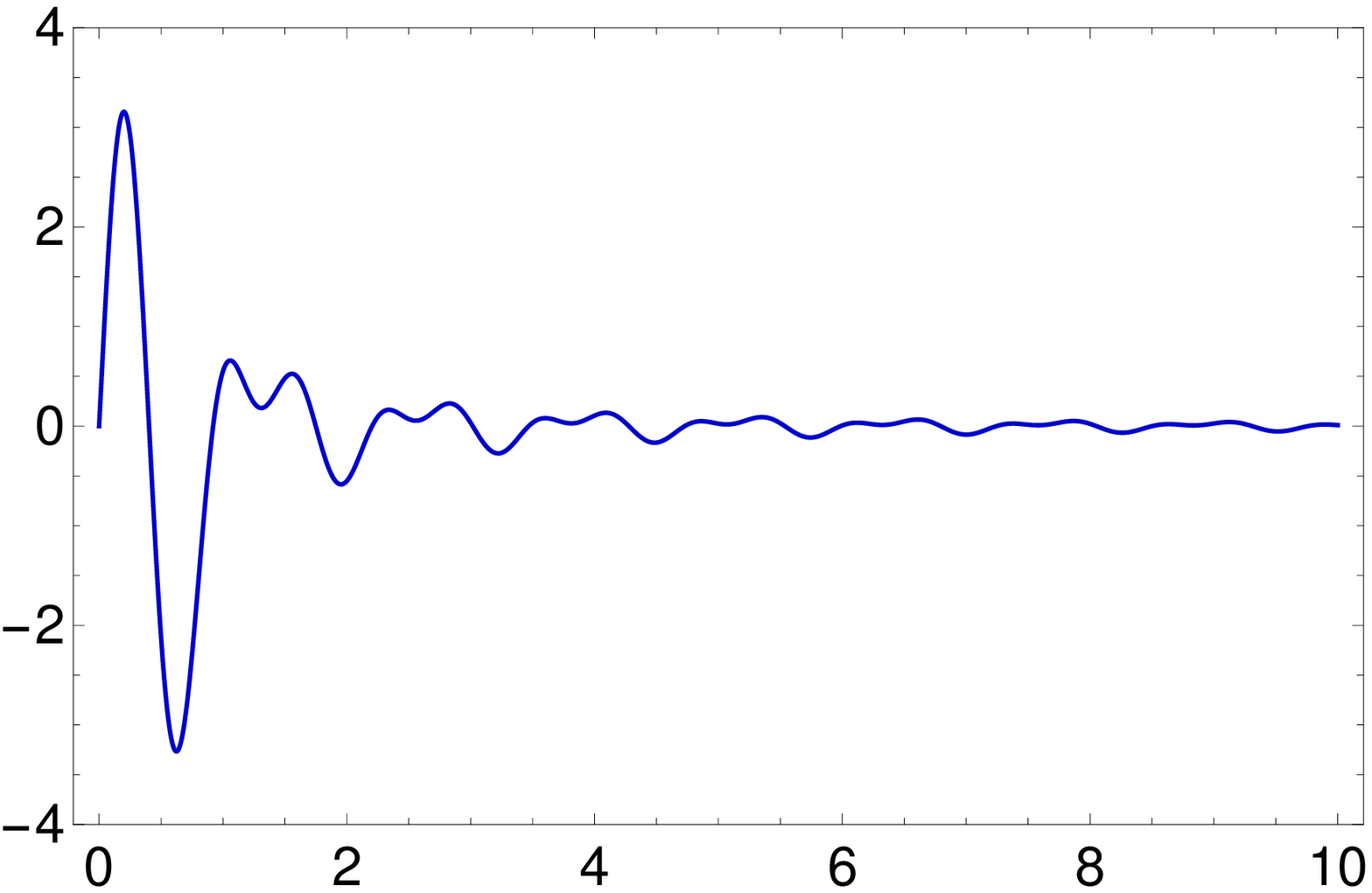,width=7.0cm}
}
\caption{Plots of the three-dimensional Green function 
for $c=1$, $a_1=0.2$, $a_2=0.1$: 
(a) $G^{L\square}_{(3)}(\bm R,\tau=1)$ for $Y=Z=0$,
(b) $G^{L\square}_{(3)}(\bm R=0,\tau)$,
(c) $\pd_\tau G^{L\square}_{(3)}(\bm R,\tau=1)$
for $Y=Z=0$,
(d) $\pd_\tau G^{L\square}_{(3)}(\bm R=0,\tau)$.}
\label{fig:3D}
\vspace*{0.2cm}
\centerline{
\epsfig{figure=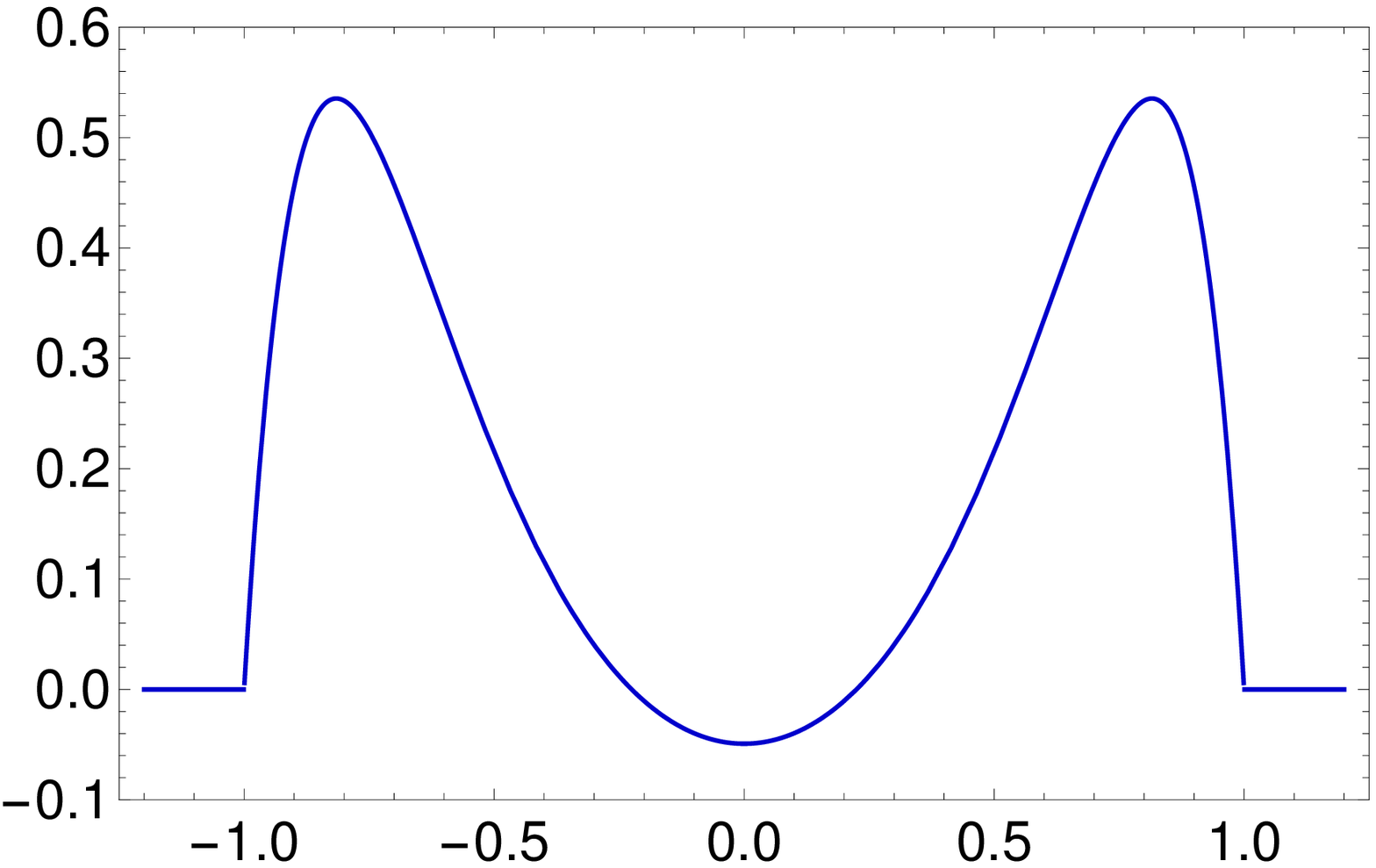,width=7.0cm}
\put(-3.5,-0.4){$X$}
\put(-7.1,-0.3){$\text{(a)}$}
\hspace*{0.4cm}
\put(-0.1,-0.3){$\text{(b)}$}
\put(3.5,-0.4){$\tau$}
\epsfig{figure=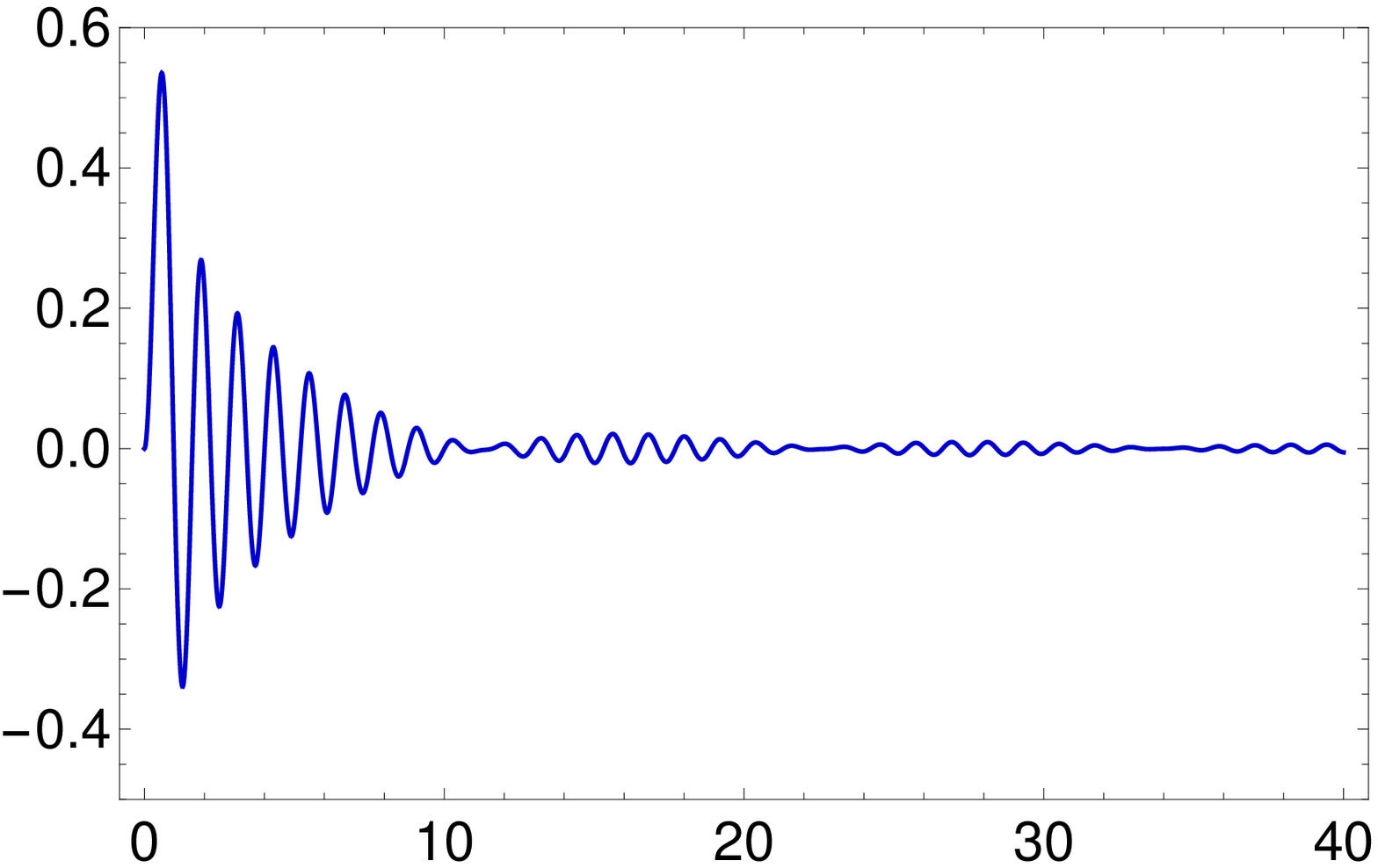,width=7.0cm}
}
\vspace*{0.2cm}
\centerline{
\epsfig{figure=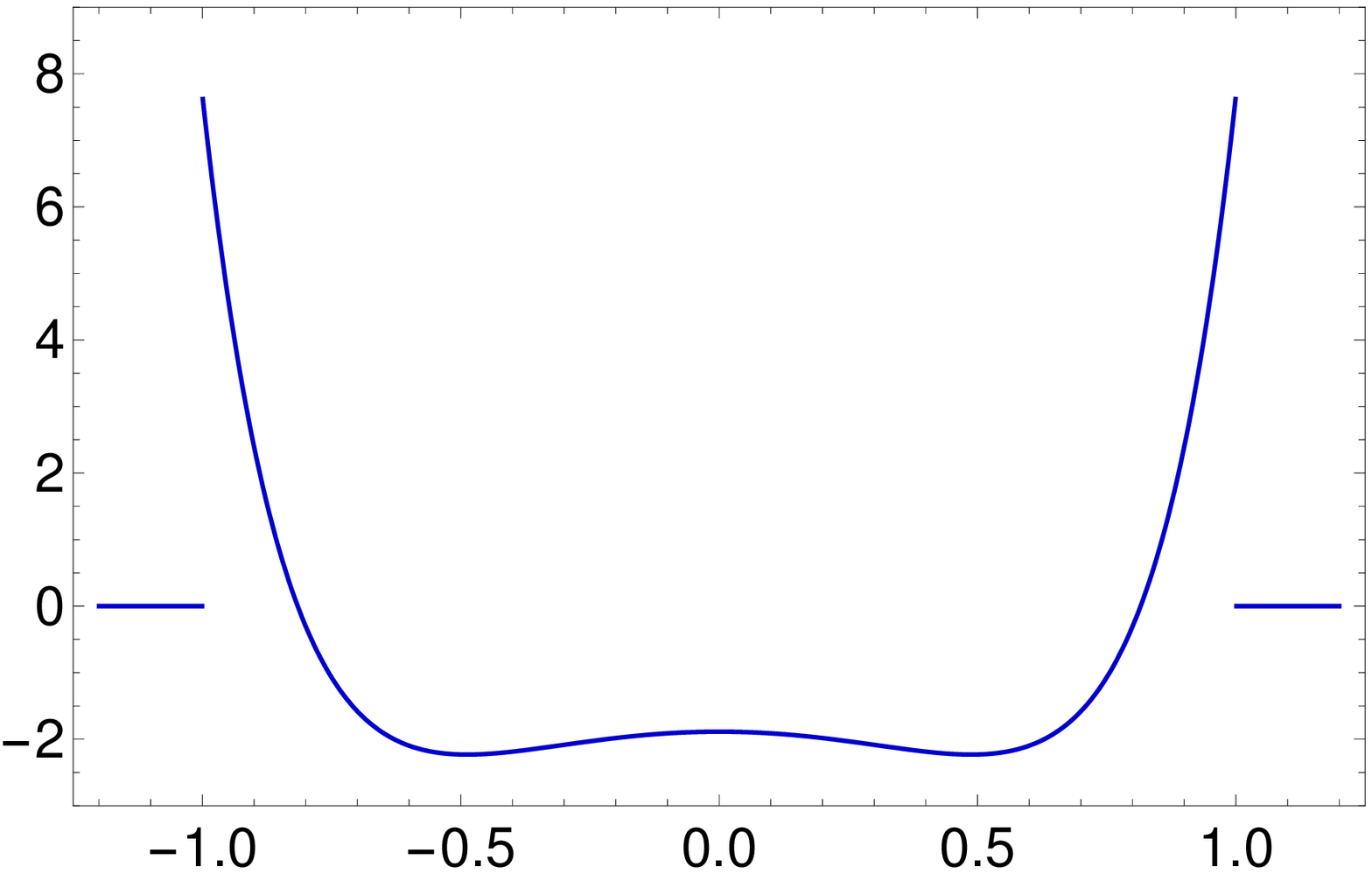,width=6.8cm}
\put(-3.5,-0.4){$X$}
\put(-7.1,-0.3){$\text{(c)}$}
\hspace*{0.4cm}
\put(3.5,-0.4){$\tau$}
\put(-0.1,-0.3){$\text{(d)}$}
\epsfig{figure=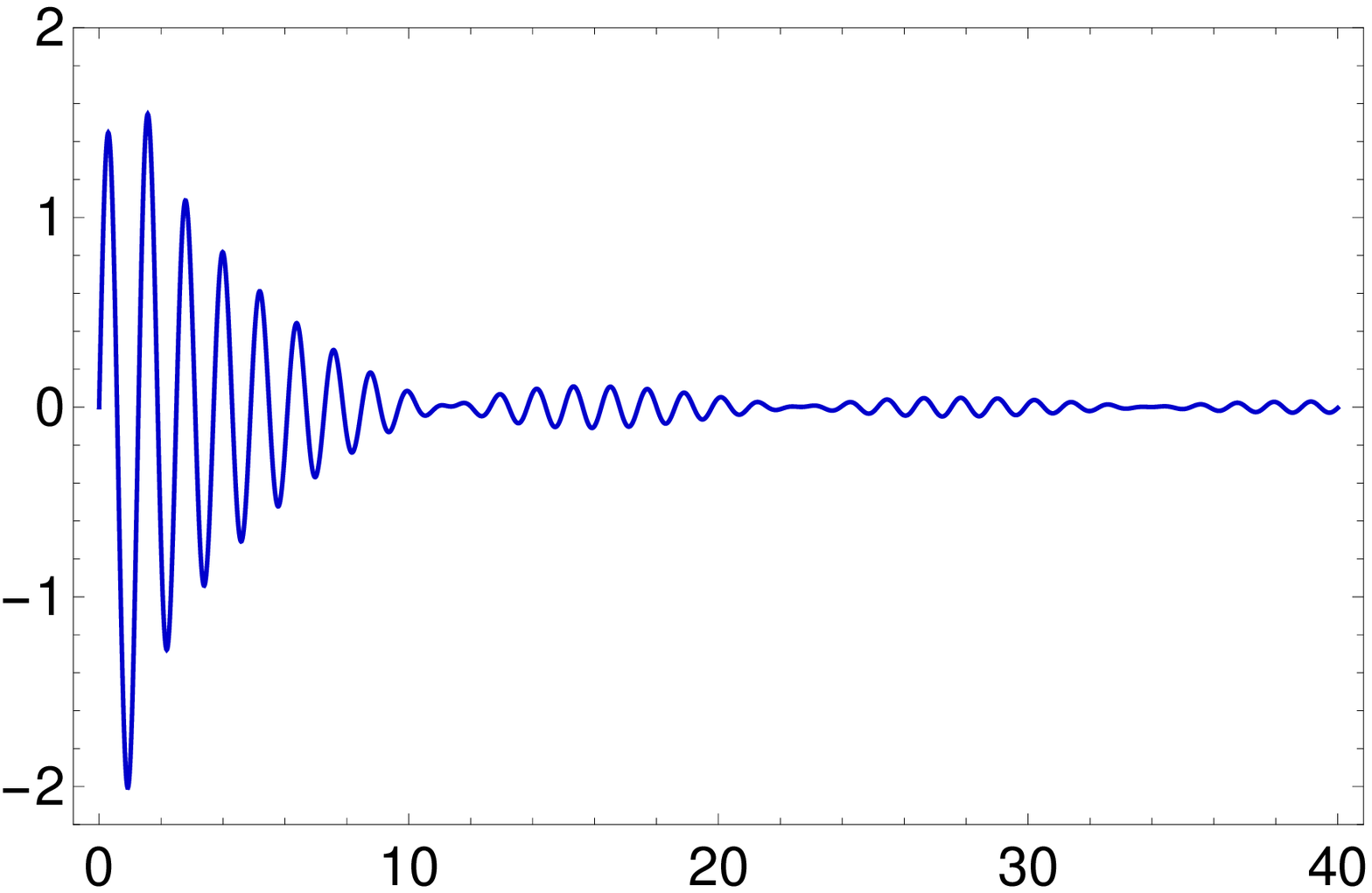,width=7.0cm}
}
\caption{Plots of the three-dimensional Green function 
for $c=1$, $a_1=0.2$, $a_2=0.18$: 
(a) $G^{L\square}_{(3)}(\bm R,\tau=1)$ for $Y=Z=0$,
(b) $G^{L\square}_{(3)}(\bm R=0,\tau)$,
(c) $\pd_\tau G^{L\square}_{(3)}(\bm R,\tau=1)$
for $Y=Z=0$,
(d) $\pd_\tau G^{L\square}_{(3)}(\bm R=0,\tau)$.}
\label{fig:3D-2}
\end{figure}

Now, taking into account that
\begin{align}
\label{rel-J1}
\lim_{z \to 0} \,\frac{1}{z}\,J_1(z)=\frac{1}{2}\,,
\end{align}
we obtain that the values of 
the Green functions~\eqref{G-BKG-3d} and \eqref{G-BP-3d}
 are zero on the light cone $c^2\tau^2-R^2=0$ 
(see Figs.~\ref{fig:3D}a and \ref{fig:3D-2}a).
Moreover, the Green functions~\eqref{G-BKG-3d} and \eqref{G-BP-3d} 
show a decreasing oscillation 
(see Fig.~\ref{fig:3D}b) and do not have a $\delta$-singularity 
unlike the Green function~\eqref{G-w-3d}.

\subsection{2D Green functions}

\begin{figure}[tp!]\unitlength1cm
\vspace*{-0.5cm}
\centerline{
\epsfig{figure=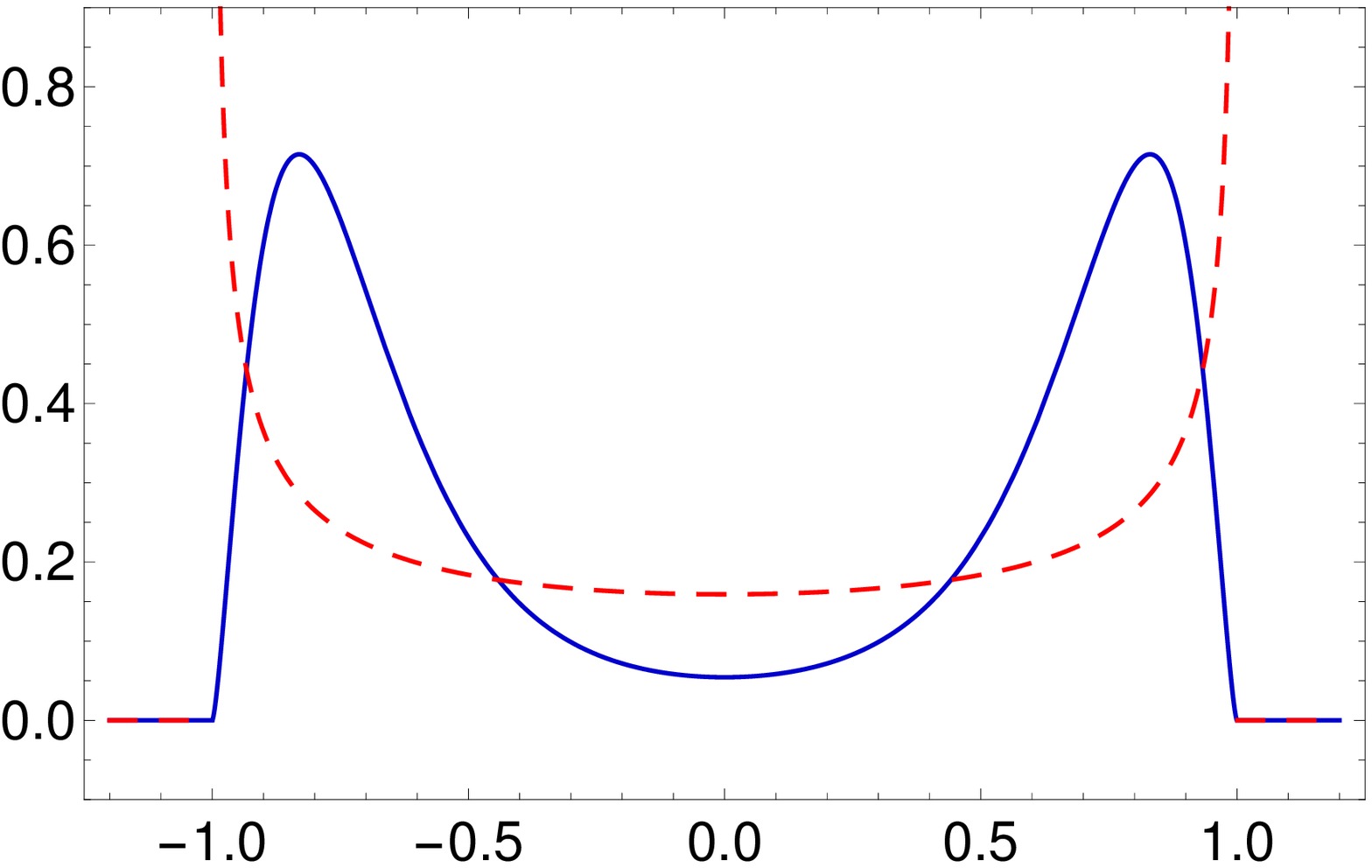,width=7.0cm}
\put(-3.5,-0.4){$X$}
\put(-7.1,-0.3){$\text{(a)}$}
\hspace*{0.4cm}
\put(-0.1,-0.3){$\text{(b)}$}
\put(3.5,-0.4){$\tau$}
\epsfig{figure=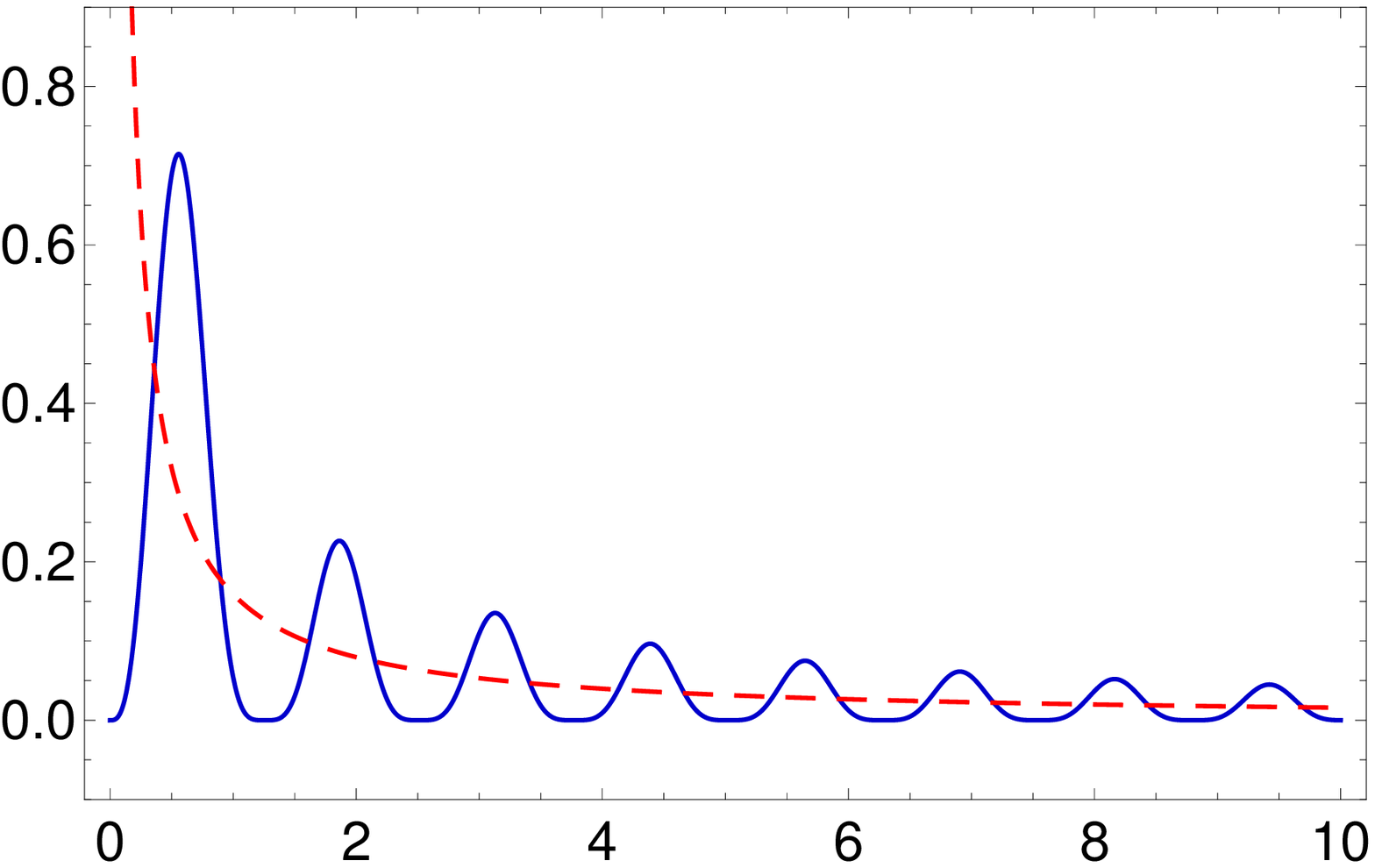,width=7.0cm}
}
\vspace*{0.2cm}
\centerline{
\epsfig{figure=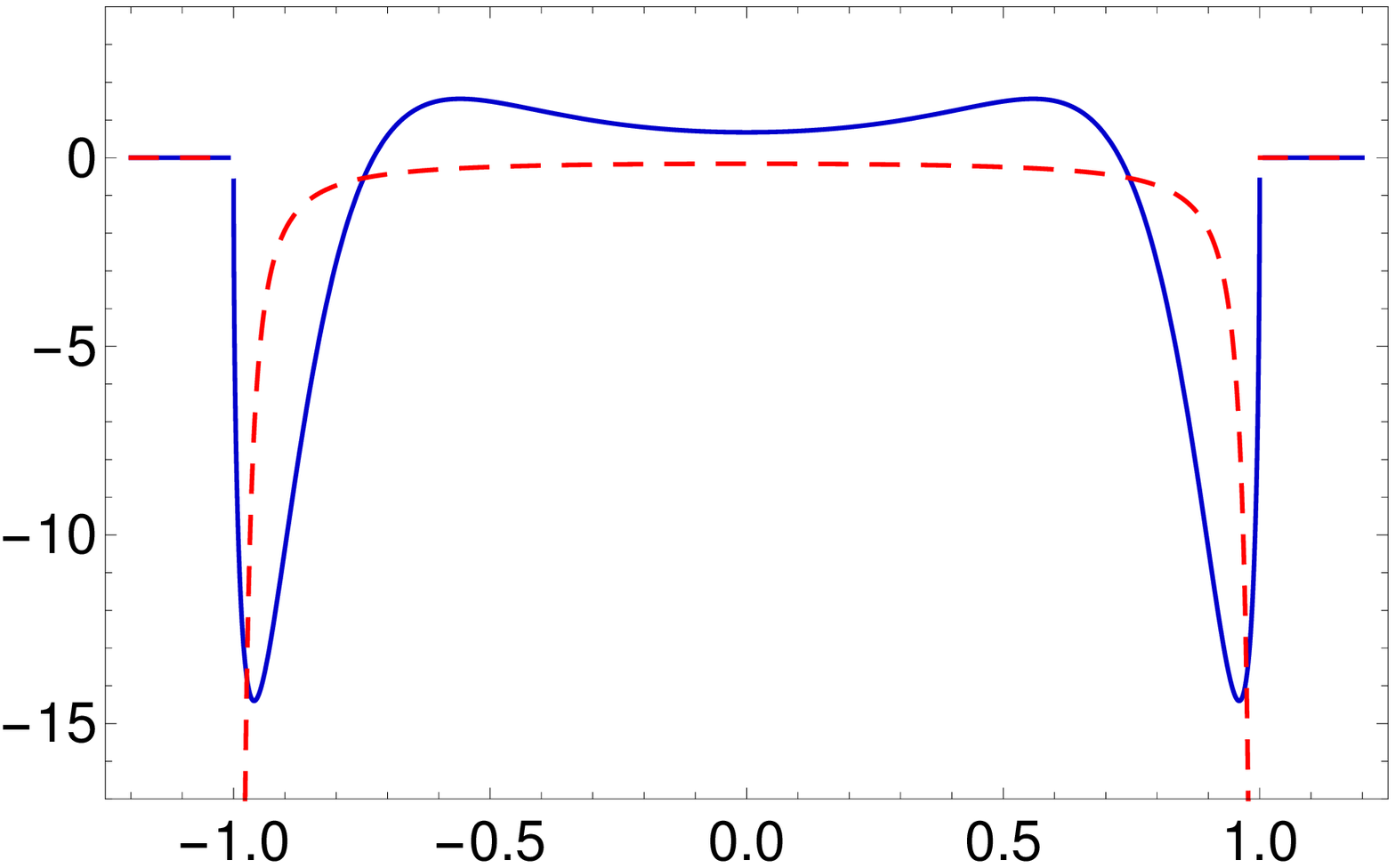,width=6.8cm}
\put(-3.5,-0.4){$X$}
\put(-7.1,-0.3){$\text{(c)}$}
\hspace*{0.4cm}
\put(3.5,-0.4){$\tau$}
\put(-0.1,-0.3){$\text{(d)}$}
\epsfig{figure=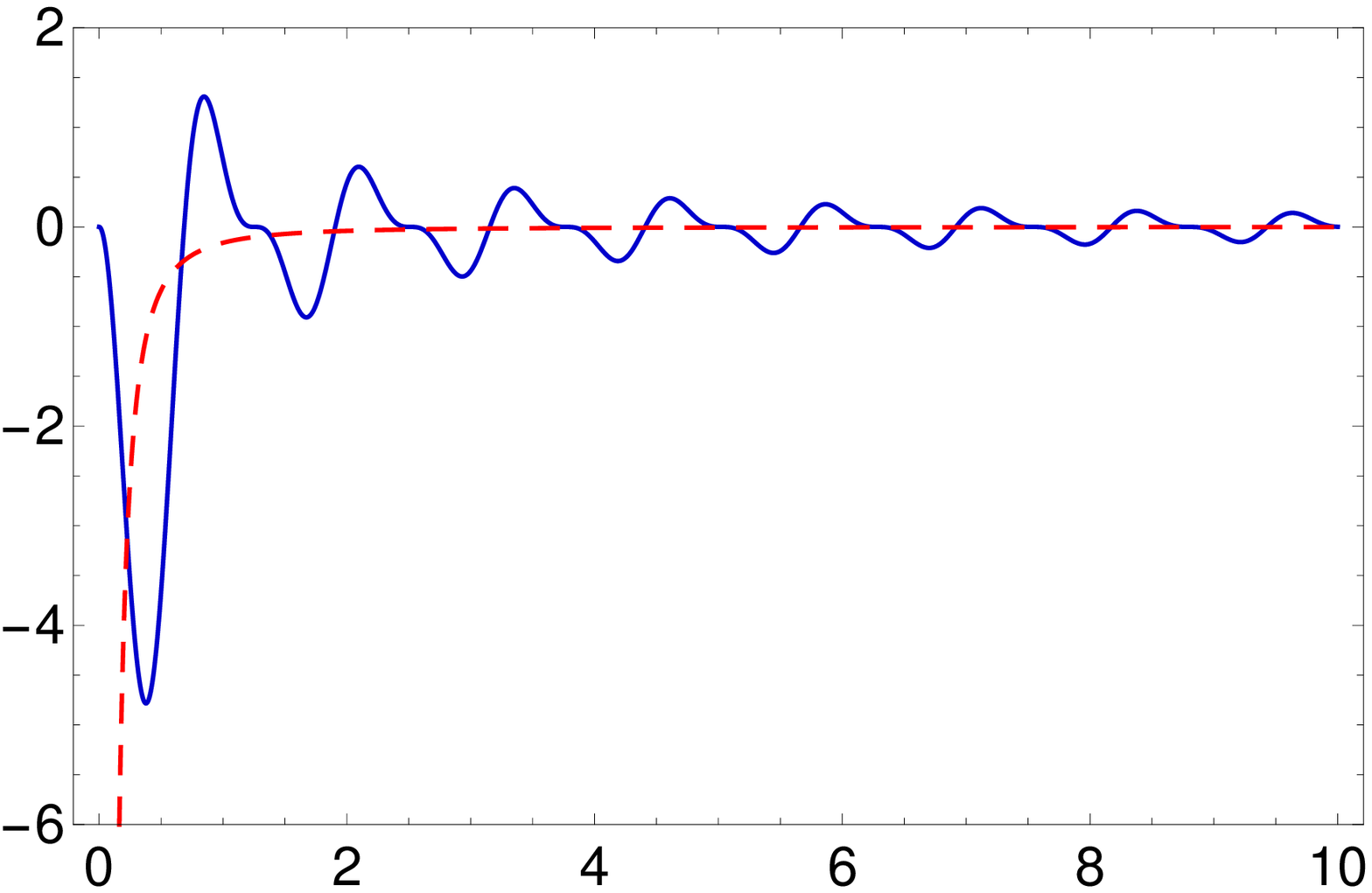,width=7.0cm}
}
\caption{Plots of the two-dimensional Green function 
for $c=1$, $a_1=0.2$, $a_2=0.1$: 
(a) $G^{L\square}_{(2)}(\bm R,\tau=1)$ for $Y=0$,
(b) $G^{L\square}_{(2)}(\bm R=0,\tau)$,
(c) $\pd_\tau G^{L\square}_{(2)}(\bm R,\tau=1)$ for $Y=0$,
(d) $\pd_\tau G^{L\square}_{(2)}(\bm R=0,\tau)$
(red dashed curves are the classical Green function $G_{(2)}^\square$
and  $\pd_\tau G_{(2)}^\square$).}
\label{fig:2D}
\vspace*{0.2cm}
\centerline{
\epsfig{figure=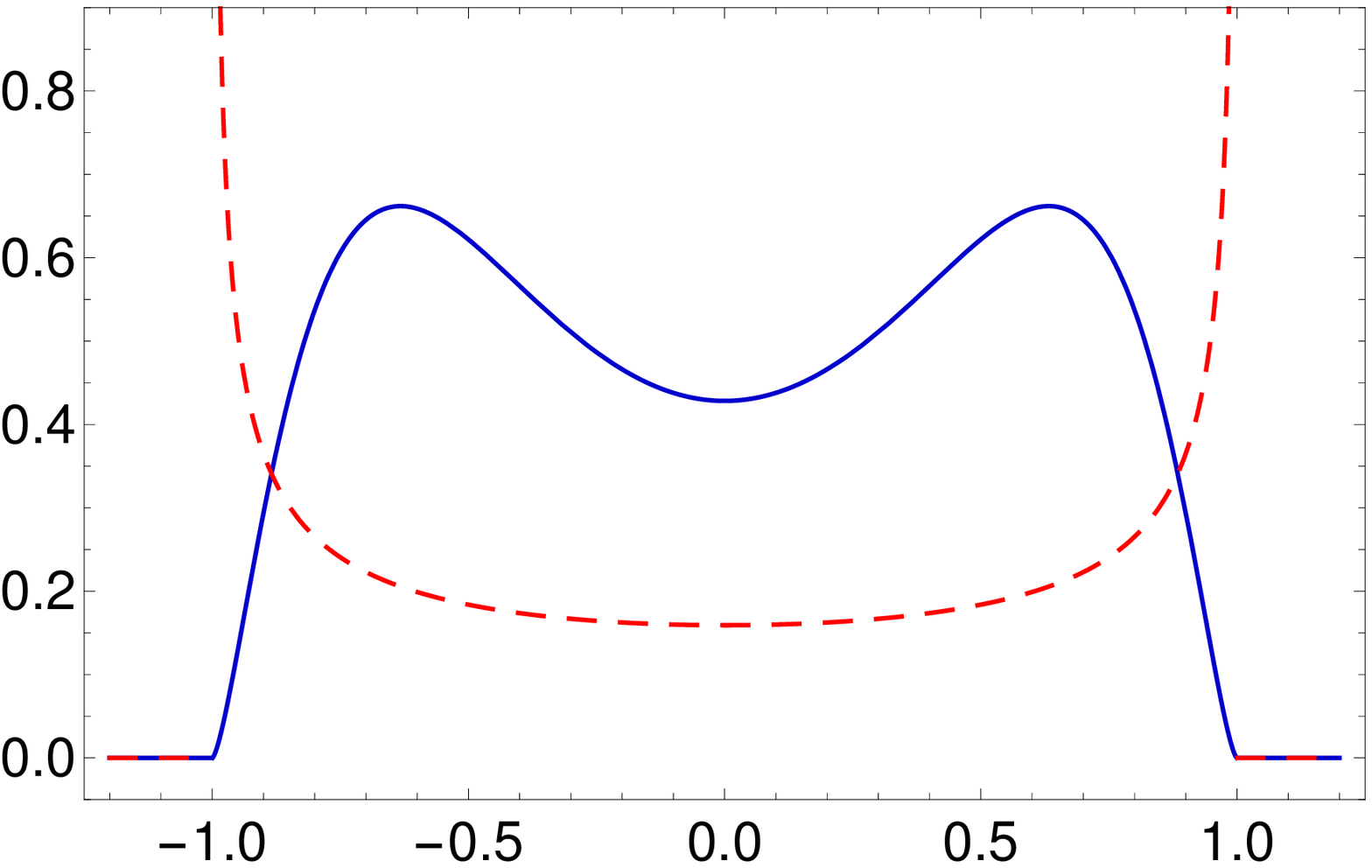,width=7.0cm}
\put(-3.5,-0.4){$X$}
\put(-7.1,-0.3){$\text{(a)}$}
\hspace*{0.4cm}
\put(-0.1,-0.3){$\text{(b)}$}
\put(3.5,-0.4){$\tau$}
\epsfig{figure=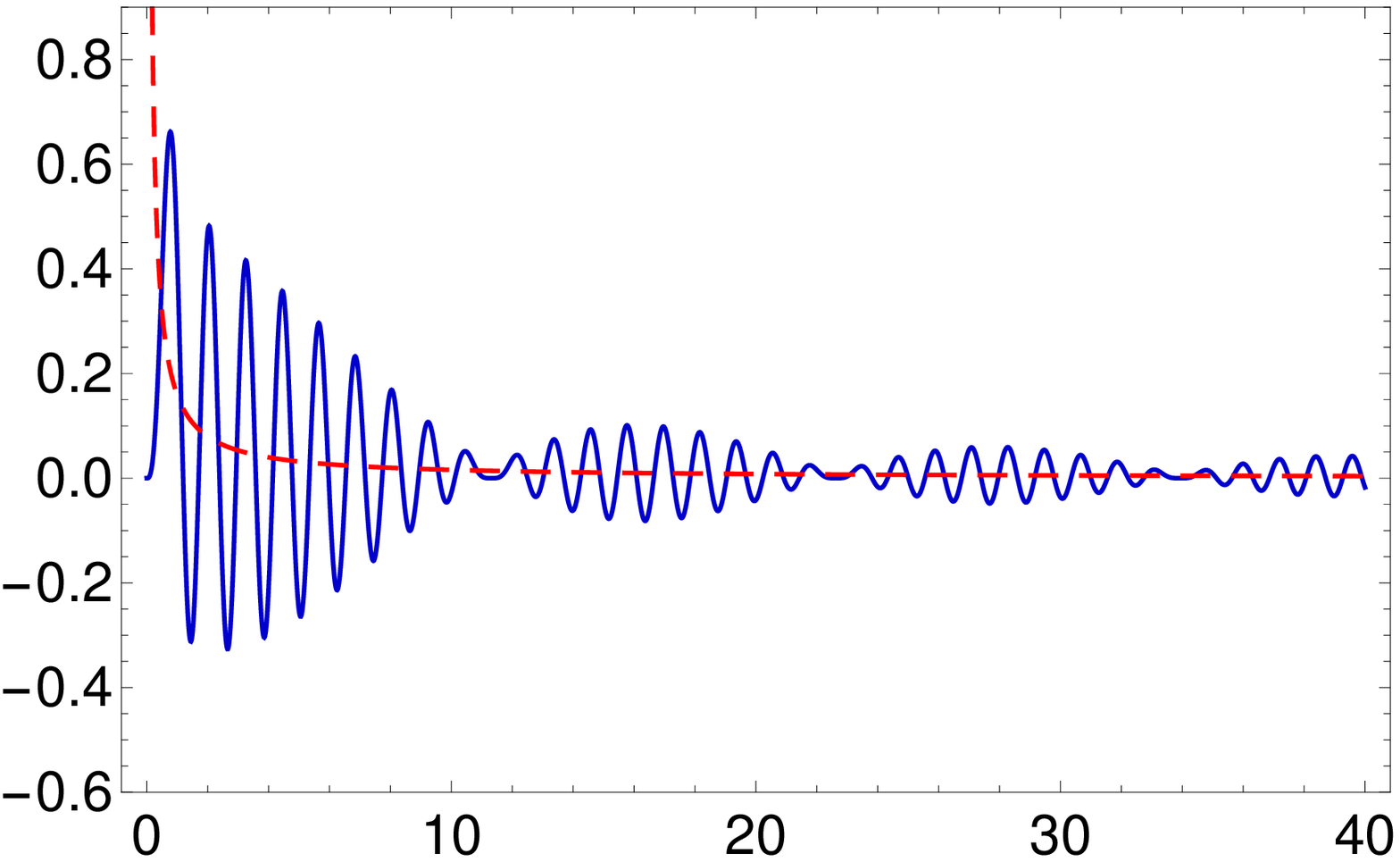,width=7.0cm}
}
\vspace*{0.2cm}
\centerline{
\epsfig{figure=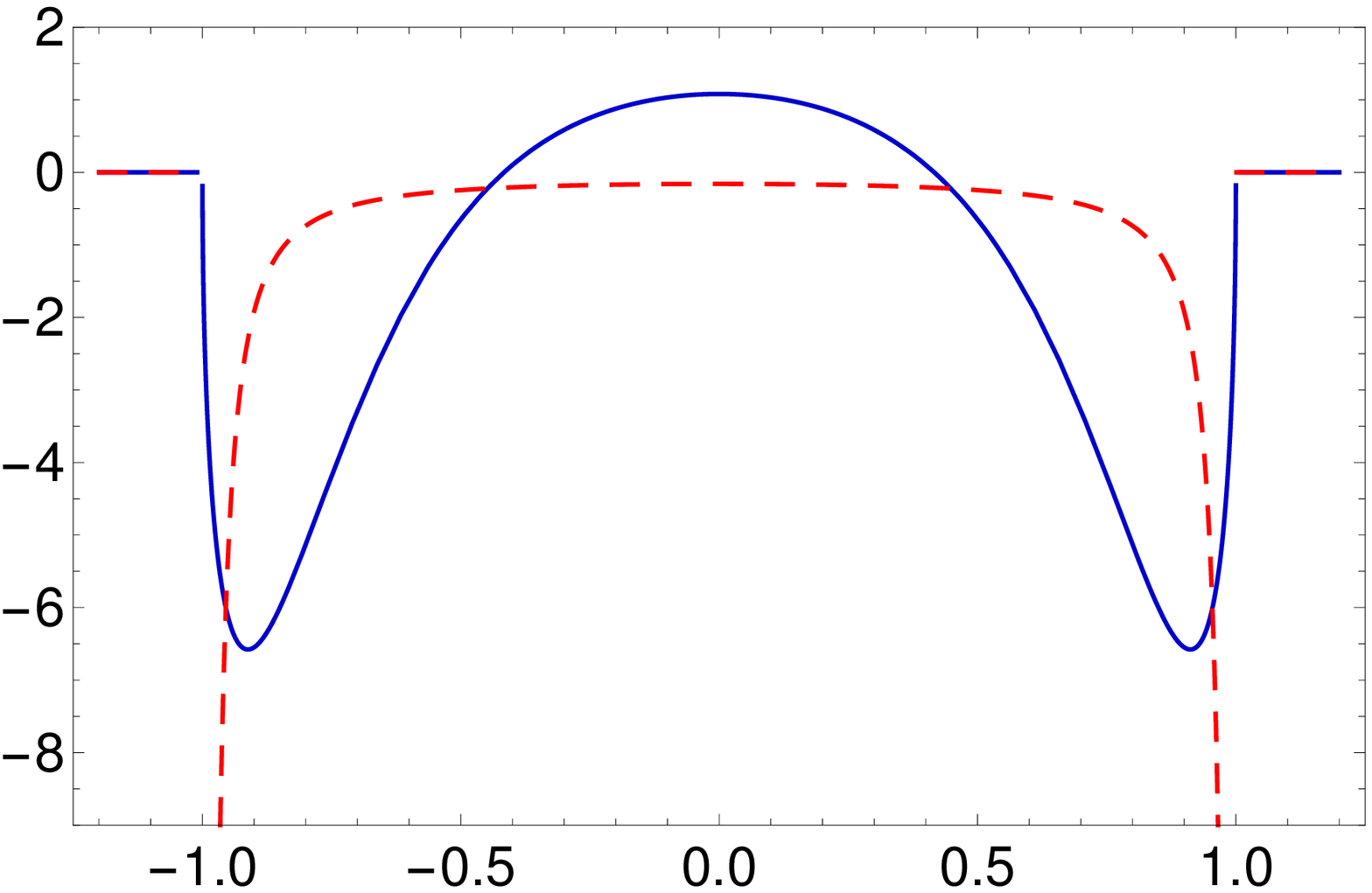,width=6.8cm}
\put(-3.5,-0.4){$X$}
\put(-7.1,-0.3){$\text{(c)}$}
\hspace*{0.4cm}
\put(3.5,-0.4){$\tau$}
\put(-0.1,-0.3){$\text{(d)}$}
\epsfig{figure=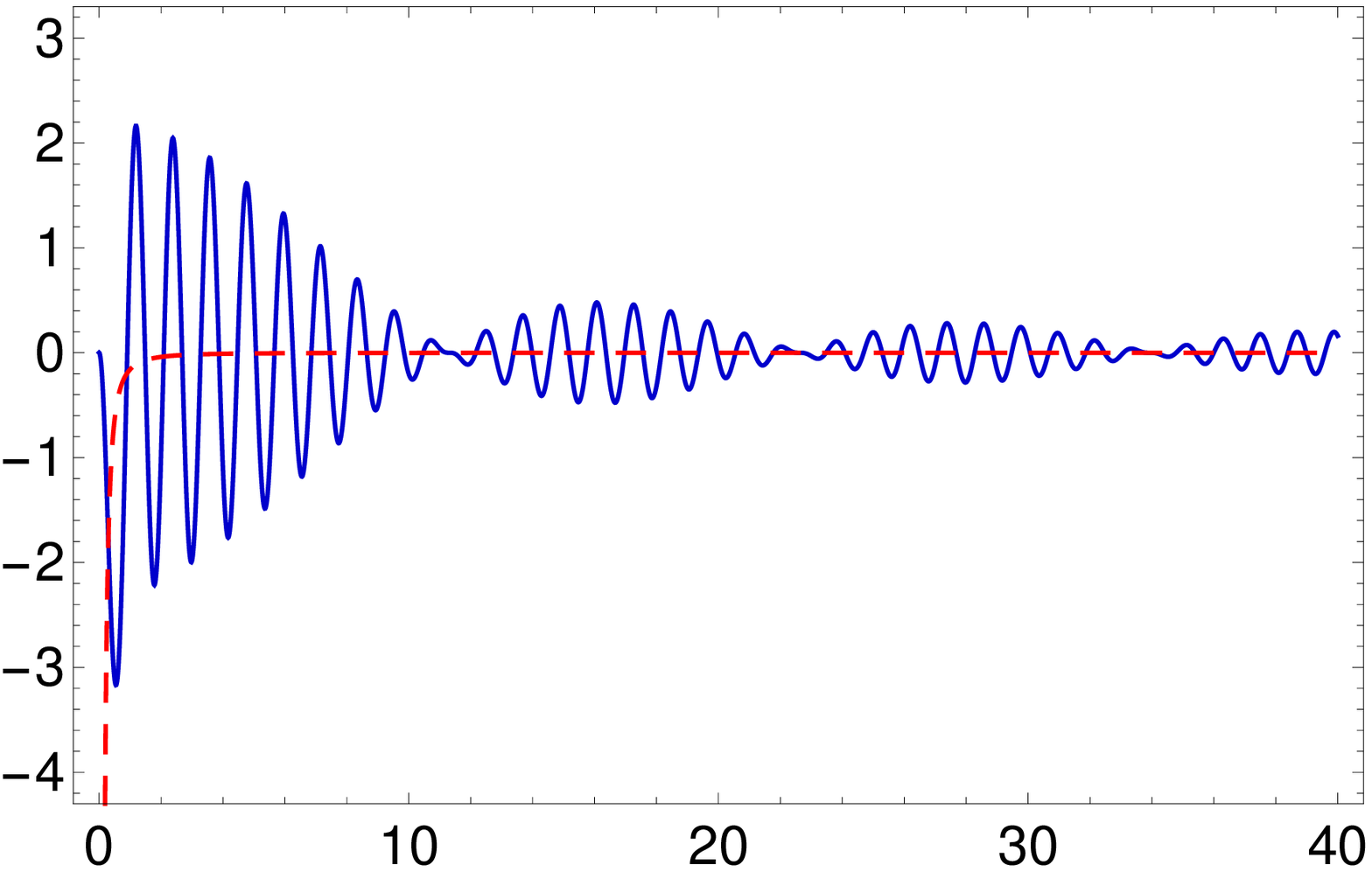,width=7.0cm}
}
\caption{Plots of the two-dimensional Green function 
for $c=1$, $a_1=0.2$, $a_2=0.18$: 
(a) $G^{L\square}_{(2)}(\bm R,\tau=1)$ for $Y=0$,
(b) $G^{L\square}_{(2)}(\bm R=0,\tau)$,
(c) $\pd_\tau G^{L\square}_{(2)}(\bm R,\tau=1)$ for $Y=0$,
(d) $\pd_\tau G^{L\square}_{(2)}(\bm R=0,\tau)$
(red dashed curves are the classical Green function $G_{(2)}^\square$
and  $\pd_\tau G_{(2)}^\square$).}
\label{fig:2D-2}
\end{figure}

The two-dimensional retarded Green functions 
of the d'Alembert operator~\eqref{wave}, the Klein-Gordon operator,
the bi-Klein-Gordon operator~\eqref{KGE} and the bi-Klein-Gordon-d'Alembert operator are given by
\begin{align}
\label{G-w-2d}
G_{(2)}^\square(\bm R, \tau)&=\frac{c}{2\pi}\,
\frac{H\big(c\tau-R\big)}{\sqrt{c^2\tau^2-R^2}}\,,\\
\label{G-KG-2d}
G_{(2)}^{\rm KG}(\bm R, \tau)
&=\frac{c}{2\pi a_1^2}\,
\frac{H\big(c\tau-R\big)}{\sqrt{c^2 \tau^2-R^2}}\, 
\cos\bigg( \frac{\sqrt{c^2 \tau^2-R^2}}{a_1}\bigg)\,,\\
\label{G-BKG-2d}
G_{(2)}^{L}(\bm R, \tau)
&=\frac{c}{2\pi(a_1^2-a_2^2)}\,
\frac{H\big(c\tau-R\big)}{\sqrt{c^2 \tau^2-R^2}}\, 
\bigg[
\cos\bigg( \frac{\sqrt{c^2 \tau^2-R^2}}{a_1}\bigg)
-\cos\bigg( \frac{\sqrt{c^2 \tau^2-R^2}}{a_2}\bigg)
\bigg]
\,,\\
\label{G-BP-2d}
G_{(2)}^{L\square}(\bm R, \tau)
&=\frac{c}{2\pi}\,
\frac{H\big(c\tau-R\big)}{\sqrt{c^2\tau^2-R^2}}\, 
\bigg[1-
\frac{1}{a_1^2-a_2^2}\bigg(
a_1^2 \cos\bigg( \frac{\sqrt{c^2 \tau^2-R^2}}{a_1}\bigg)
-a_2^2 \cos\bigg( \frac{\sqrt{c^2 \tau^2-R^2}}{a_2}\bigg)\bigg)
\bigg] \,,
\end{align}
where $R=\sqrt{(x-x')^2+(y-y')^2}$.
It can be seen that Eq.~\eqref{G-BP-2d} is obtained from Eq.~\eqref{BPG} using the Green functions~\eqref{G-w-2d} and \eqref{G-KG-2d}.
Let us observe that the Green function~\eqref{G-BP-2d} is zero on
the light cone 
(see Figs.~\ref{fig:2D}a and \ref{fig:2D-2}a) taking into account the limit 
\begin{align}
\lim_{z \to 0} \,\frac{1}{z}\,\cos(z)=\frac{1}{z}\,.
\end{align}
Furthermore, it can be seen that the Green function~\eqref{G-BP-2d} 
shows a decreasing oscillation around 
the classical Green function~\eqref{G-w-2d} 
(see Fig.~\ref{fig:2D}b).

\subsection{1D Green functions}

\begin{figure}[tp!]\unitlength1cm
\vspace*{-0.5cm}
\centerline{
\epsfig{figure=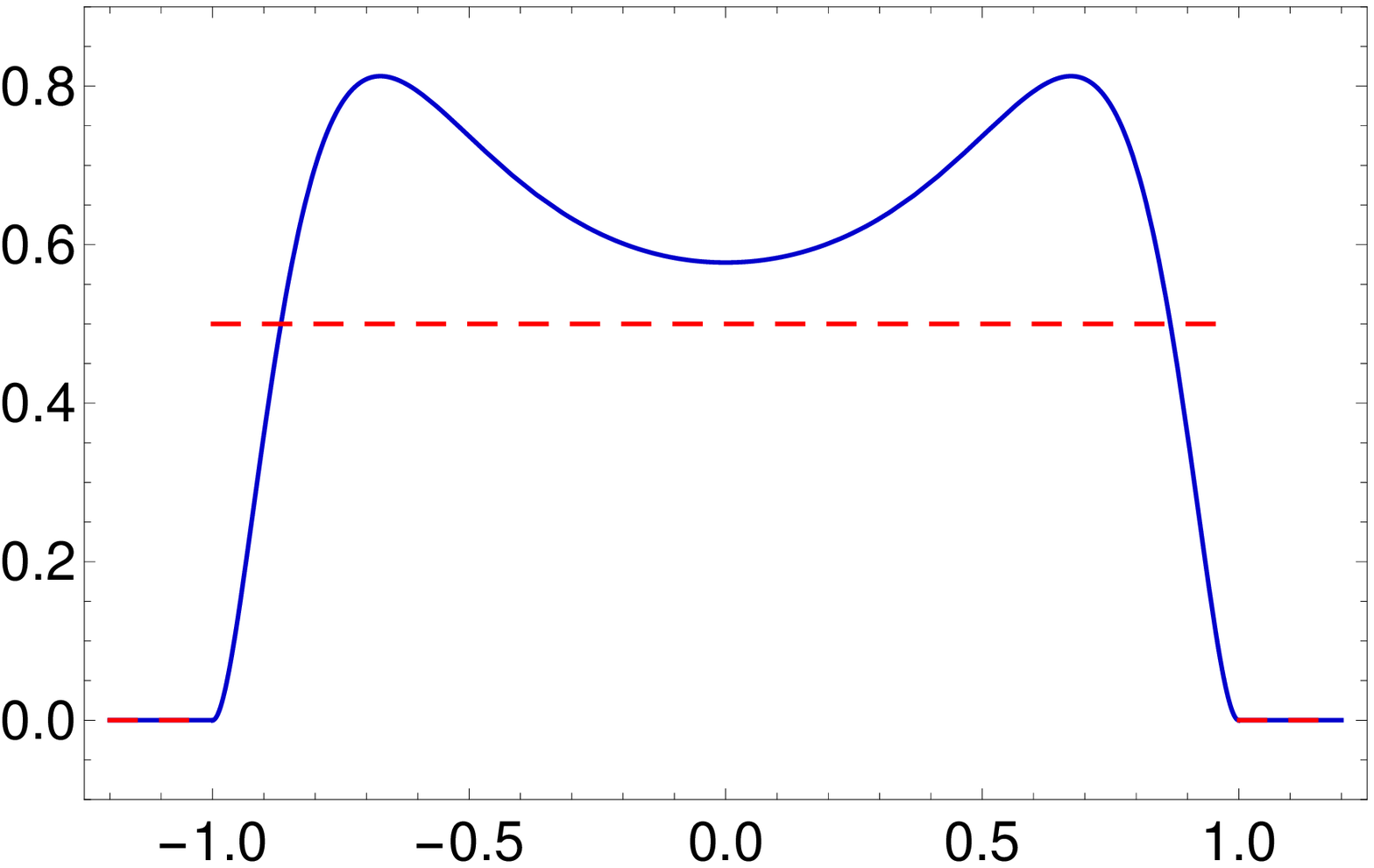,width=7.0cm}
\put(-3.5,-0.4){$X$}
\put(-7.1,-0.3){$\text{(a)}$}
\hspace*{0.4cm}
\put(-0.1,-0.3){$\text{(b)}$}
\put(3.5,-0.4){$\tau$}
\epsfig{figure=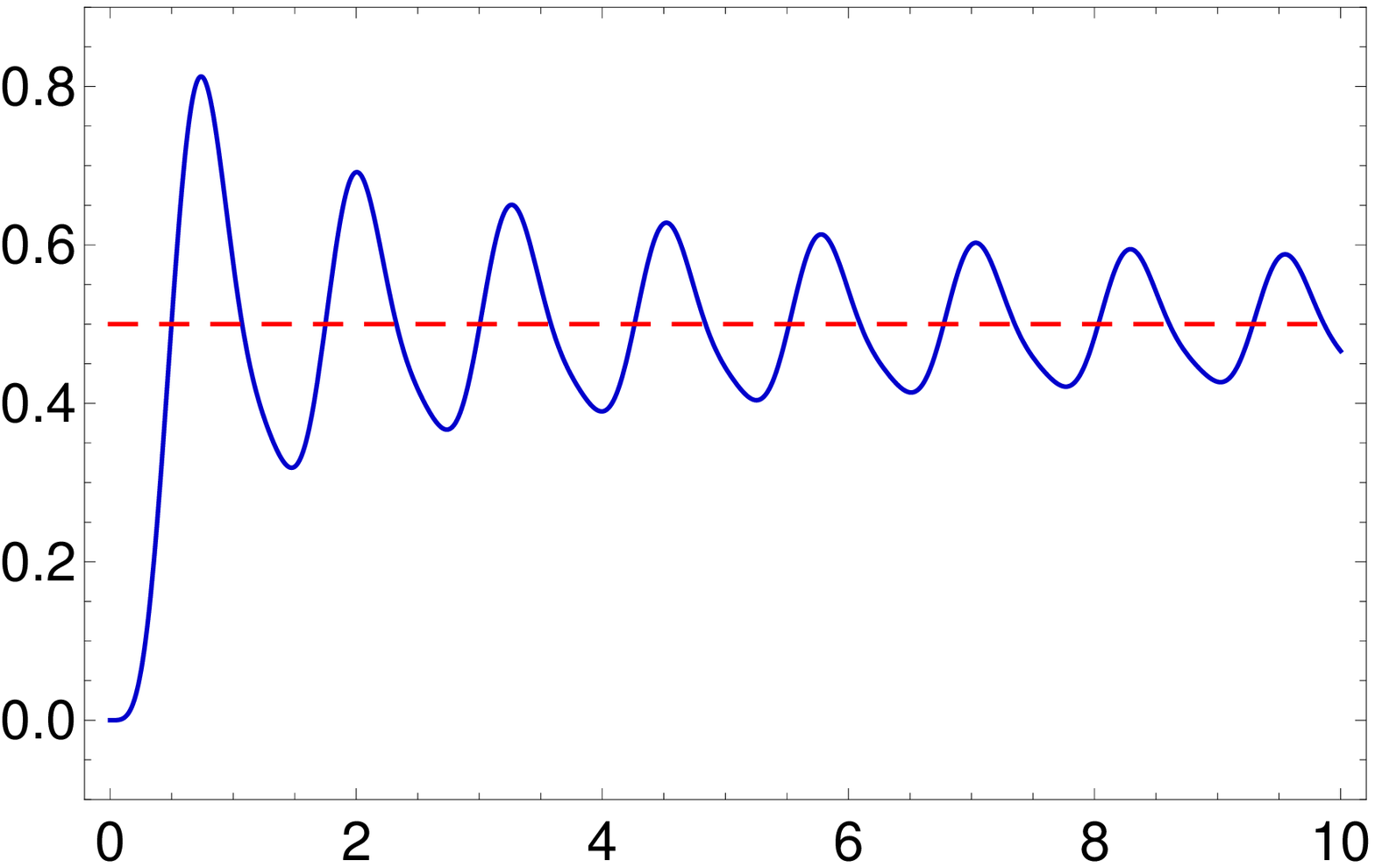,width=7.0cm}
}
\vspace*{0.2cm}
\centerline{
\epsfig{figure=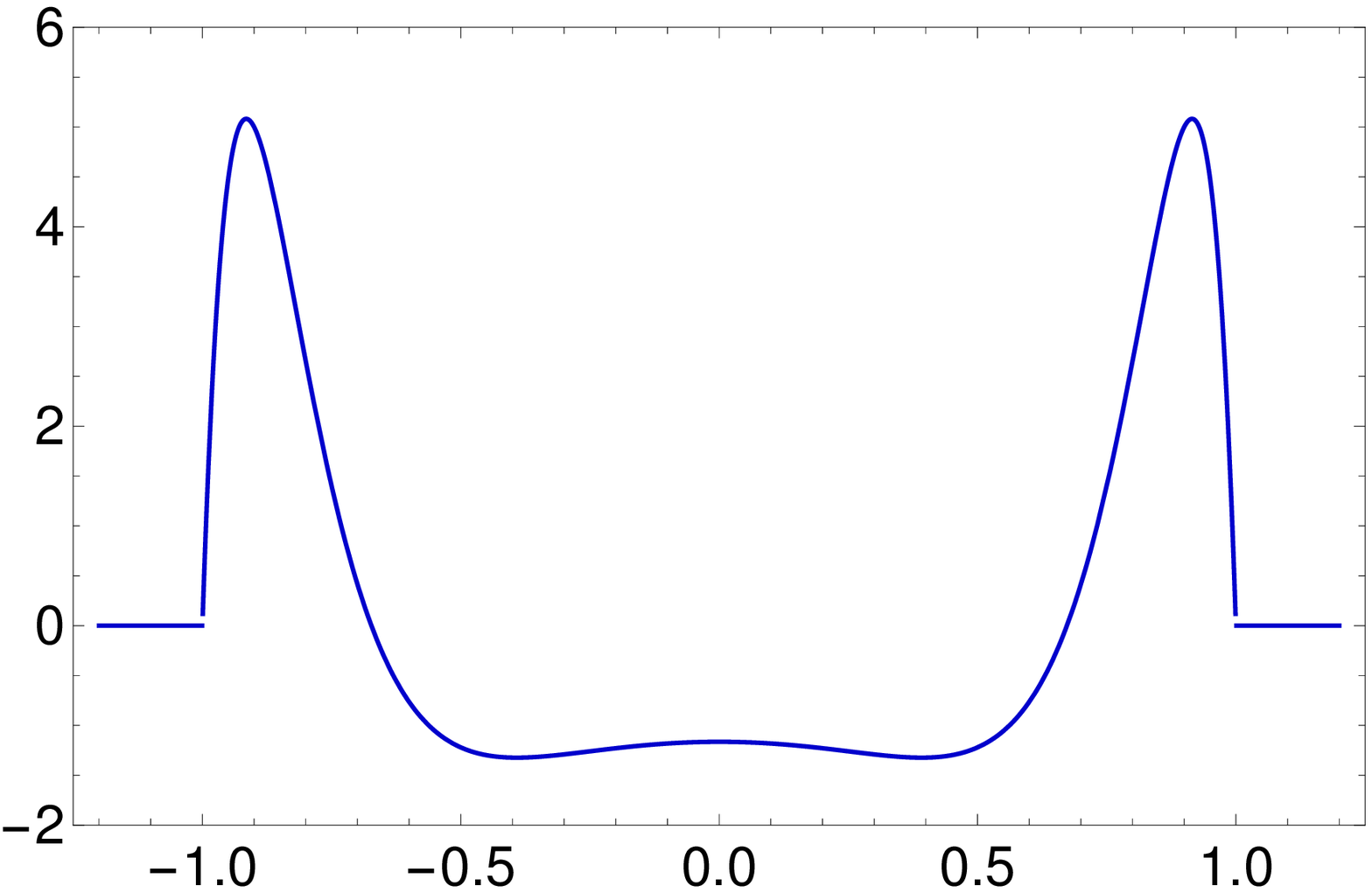,width=6.8cm}
\put(-3.5,-0.4){$X$}
\put(-7.1,-0.3){$\text{(c)}$}
\hspace*{0.4cm}
\put(3.5,-0.4){$\tau$}
\put(-0.1,-0.3){$\text{(d)}$}
\epsfig{figure=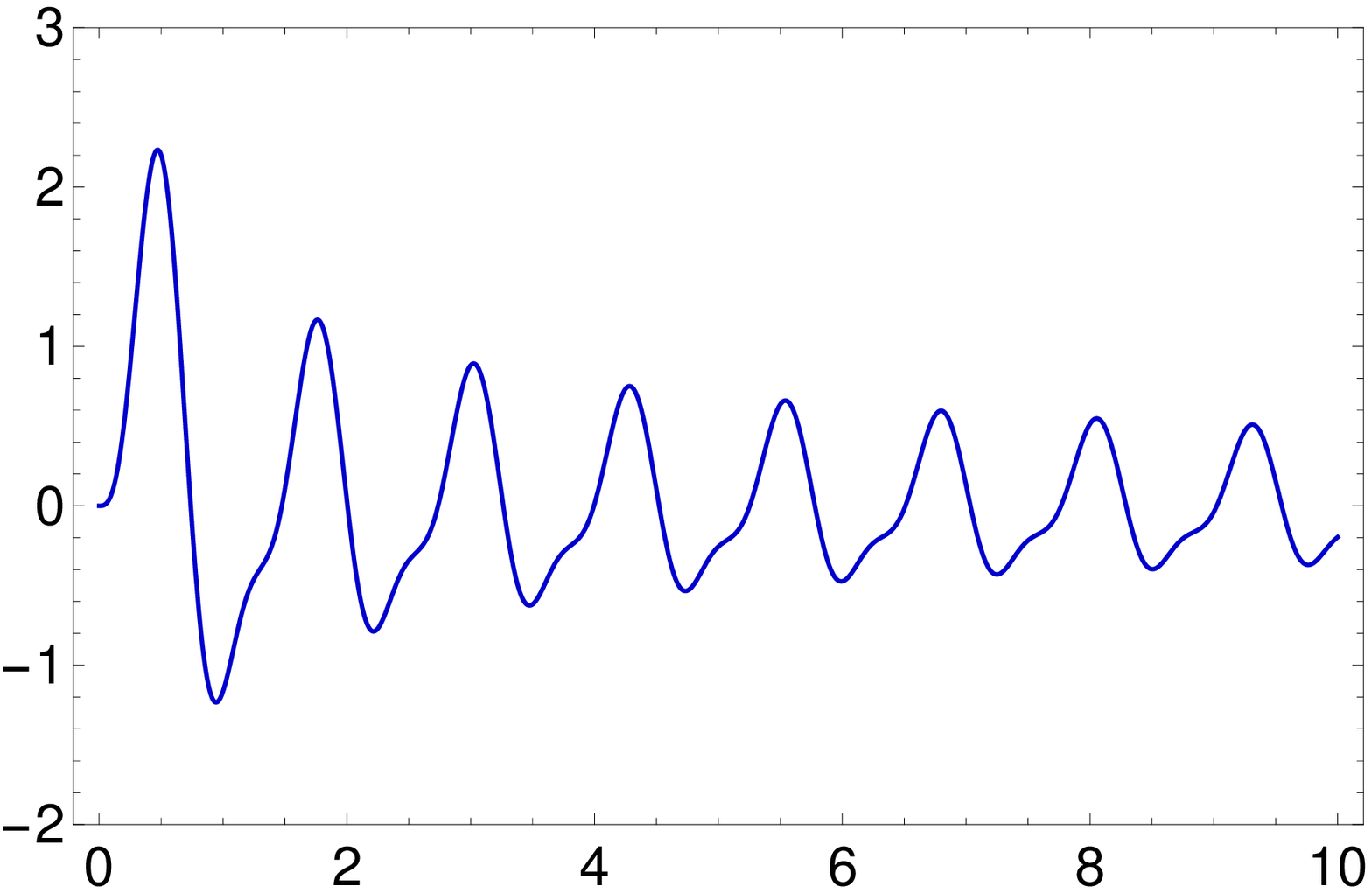,width=7.0cm}
}
\caption{Plots of the one-dimensional Green function 
for $c=1$, $a_1=0.2$, $a_2=0.1$: 
(a) $G^{L\square}_{(1)}(X,\tau=1)$,
(b) $G^{L\square}_{(1)}(X=0,\tau)$,
(c) $\pd_\tau G^{L\square}_{(1)}(X,\tau=1)$,
(d) $\pd_\tau G^{L\square}_{(1)}(X=0,\tau)$
(red dashed curves are the classical Green function $G_{(1)}^\square$).}
\label{fig:1D}
\vspace*{0.2cm}
\centerline{
\epsfig{figure=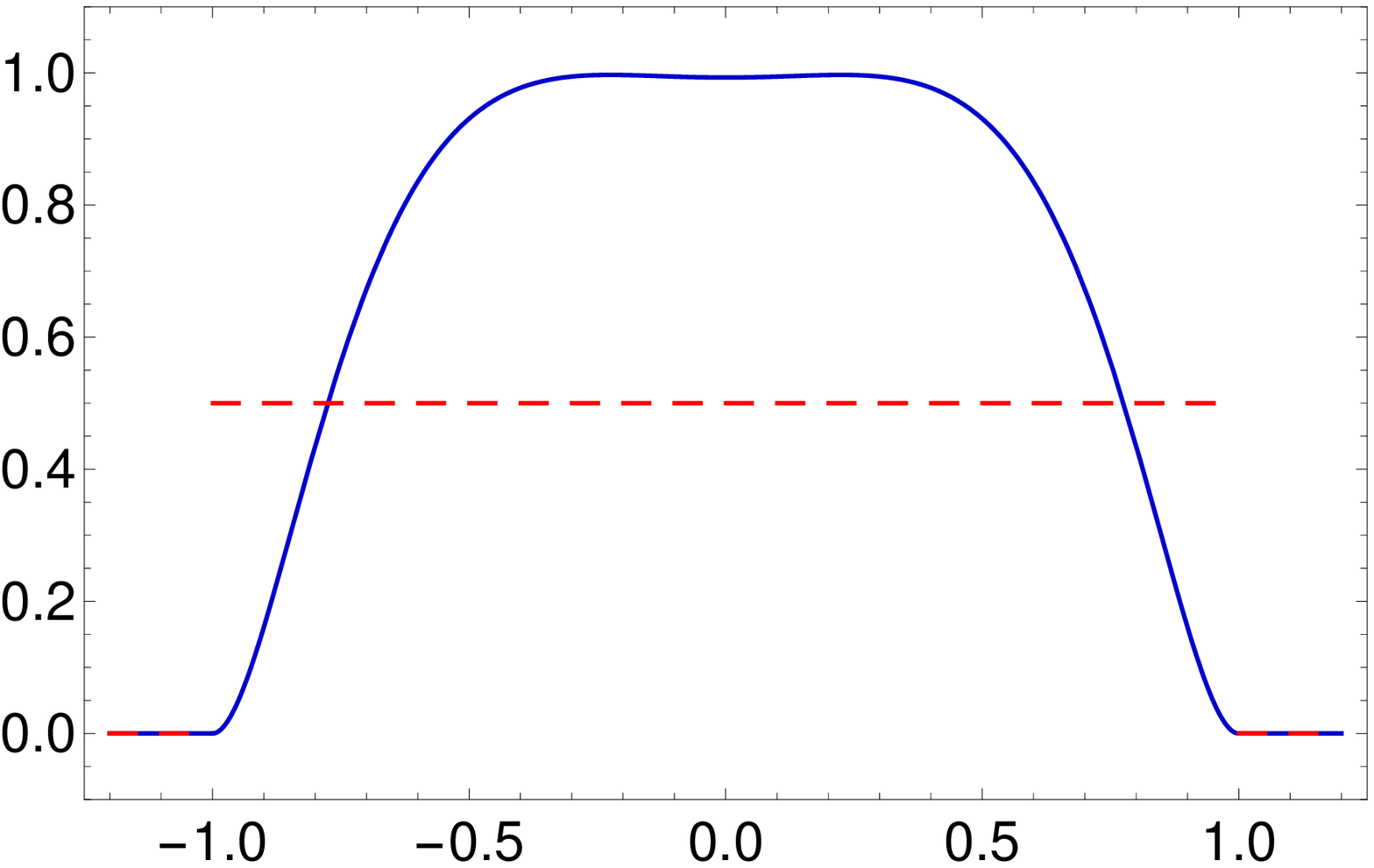,width=7.0cm}
\put(-3.5,-0.4){$X$}
\put(-7.1,-0.3){$\text{(a)}$}
\hspace*{0.4cm}
\put(-0.1,-0.3){$\text{(b)}$}
\put(3.5,-0.4){$\tau$}
\epsfig{figure=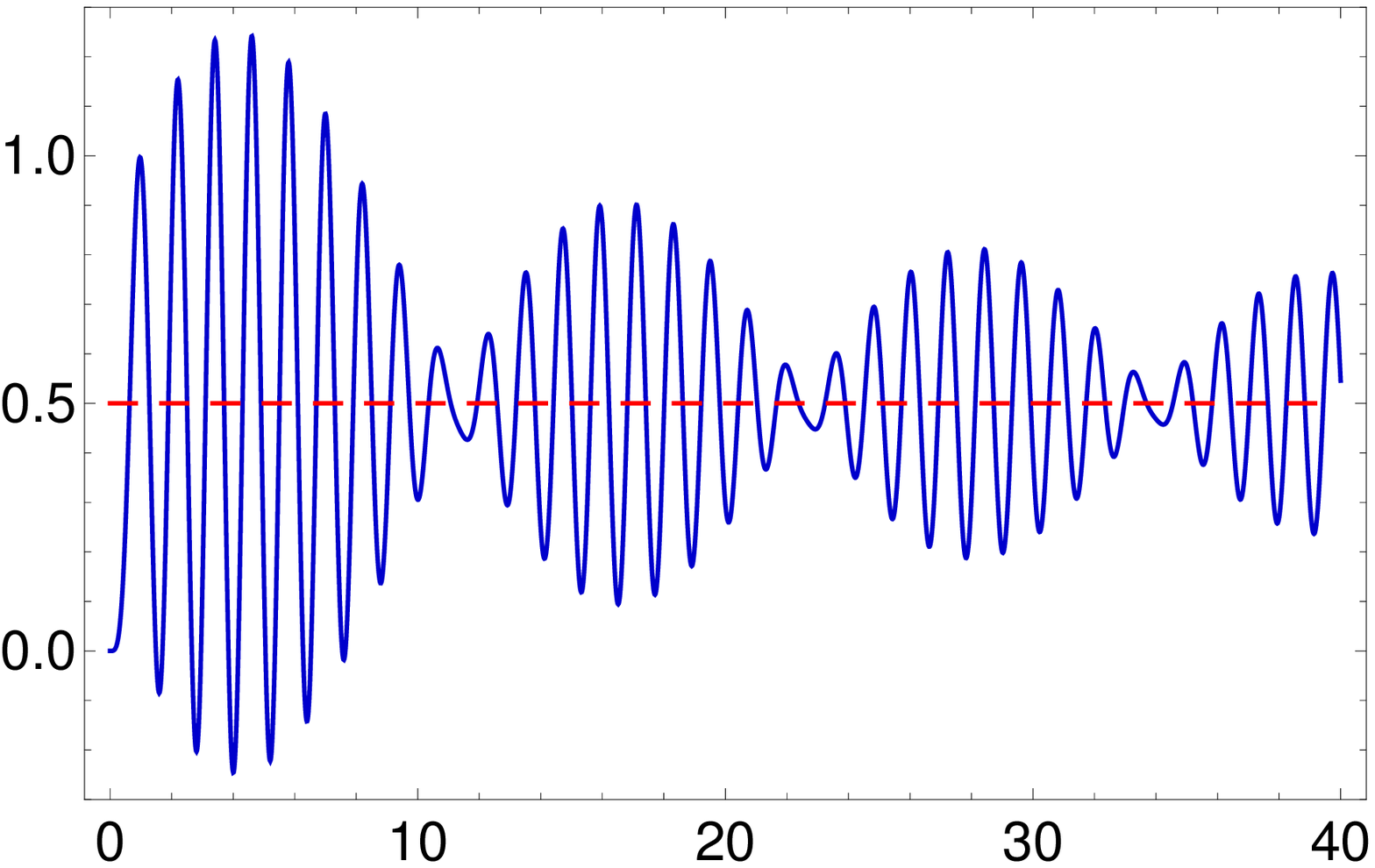,width=7.0cm}
}
\vspace*{0.2cm}
\centerline{
\epsfig{figure=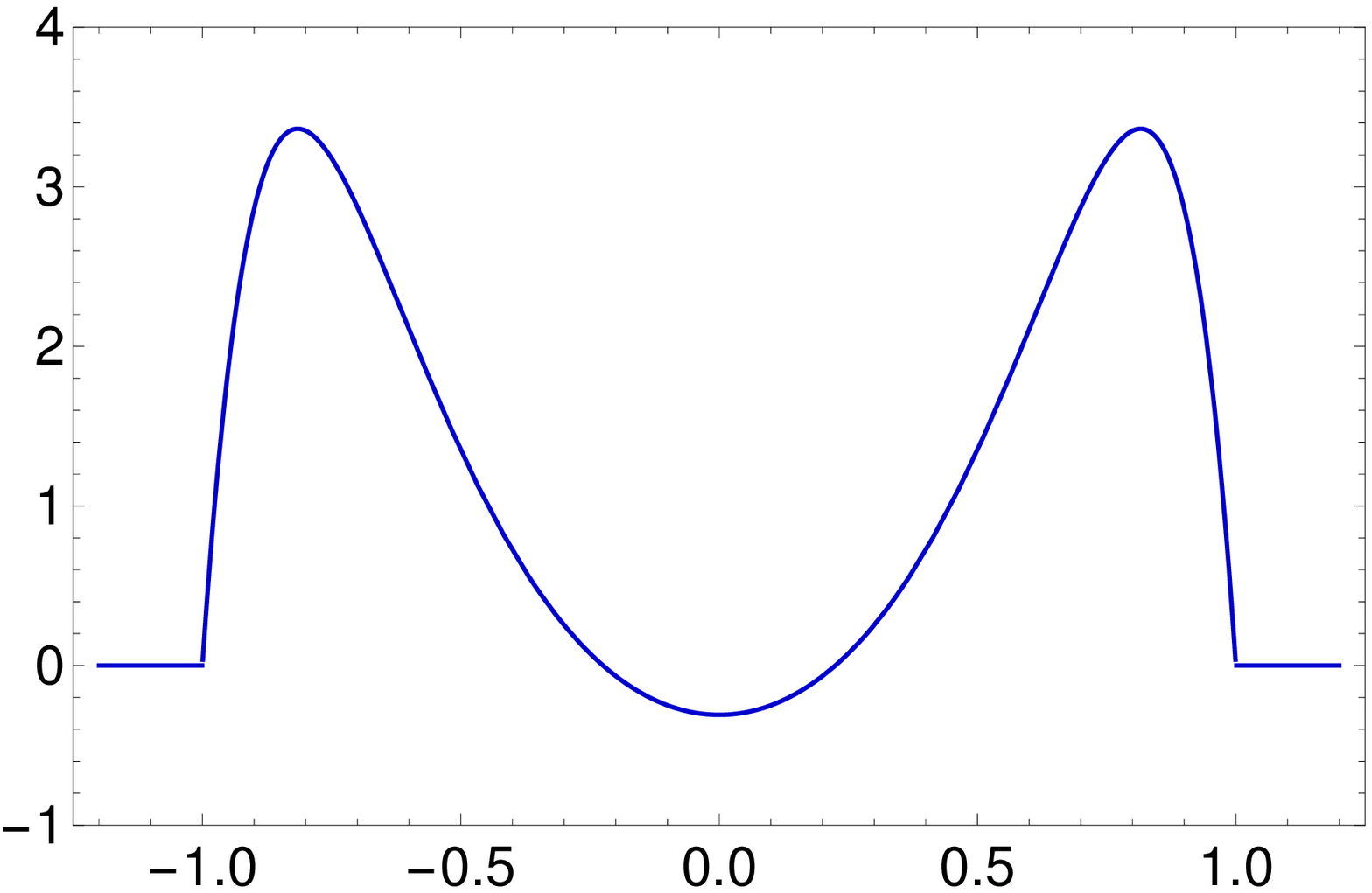,width=6.8cm}
\put(-3.5,-0.4){$X$}
\put(-7.1,-0.3){$\text{(c)}$}
\hspace*{0.4cm}
\put(3.5,-0.4){$\tau$}
\put(-0.1,-0.3){$\text{(d)}$}
\epsfig{figure=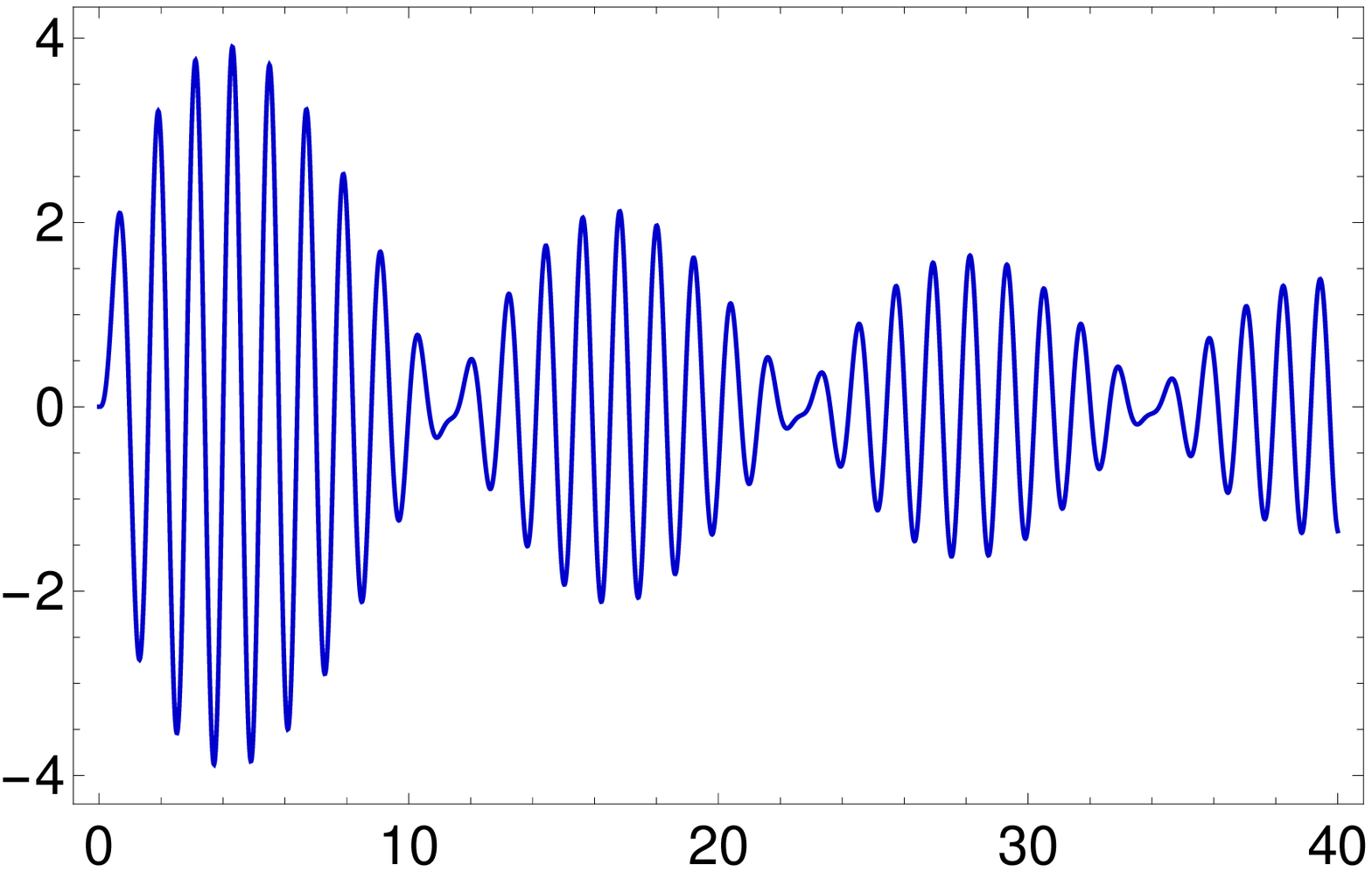,width=7.0cm}
}
\caption{Plots of the one-dimensional Green function 
for $c=1$, $a_1=0.2$, $a_2=0.18$: 
(a) $G^{L\square}_{(1)}(X,\tau=1)$,
(b) $G^{L\square}_{(1)}(X=0,\tau)$,
(c) $\pd_\tau G^{L\square}_{(1)}(X,\tau=1)$,
(d) $\pd_\tau G^{L\square}_{(1)}(X=0,\tau)$
(red dashed curves are the classical Green function $G_{(1)}^\square$).}
\label{fig:1D-2}
\end{figure}

The one-dimensional retarded Green functions of the d'Alembert operator~\eqref{wave}, the Klein-Gordon operator, 
the bi-Klein-Gordon operator~\eqref{KGE}
and the bi-Klein-Gordon-d'Alembert operator are given by
\begin{align}
\label{G-w-1d}
G_{(1)}^\square(X,\tau)&=\frac{c}{2}\, H\big(c\tau-|X|)\,,\\
\label{G-KG-1d}
G_{(1)}^{\rm KG}(X,\tau)
&=\frac{c}{2 a_1^2}\,
H\big(c\tau-|X|\big) \,
J_0 \bigg( \frac{\sqrt{c^2 \tau^2-X^2}}{a_1}\bigg)\,,\\
\label{G-BKG-1d}
G_{(1)}^{L}(X,\tau)
&=\frac{c}{2 (a_1^2-a_2^2)}\,
H\big(c\tau-|X|\big) 
\bigg[J_0 \bigg( \frac{\sqrt{c^2 \tau^2-X^2}}{a_1}\bigg)
-J_0 \bigg( \frac{\sqrt{c^2 \tau^2-X^2}}{a_2}\bigg)\bigg]
\,,\\
\label{G-BP-1d}
G_{(1)}^{L\square}(X,\tau)
&=\frac{c}{2}\, H\big(c\tau-|X|)
\bigg[1-
\frac{1}{a_1^2-a_2^2}\bigg(
a_1^2\, J_0 \bigg( \frac{\sqrt{c^2 \tau^2-X^2}}{a_1}\bigg)
-a_2^2\, J_0 \bigg( \frac{\sqrt{c^2 \tau^2-X^2}}{a_2}\bigg)
\bigg)
\bigg]\,,
\end{align}
where $X=x-x'$ and $J_0$ is the Bessel function 
of the first kind of order 0. 
Eq.~\eqref{G-BP-1d} is obtained from Eq.~\eqref{BPG} using the Green functions~\eqref{G-w-1d} and \eqref{G-KG-1d}.
Note that the Green function~\eqref{G-BP-1d} becomes zero on
the light cone 
(see Figs.~\ref{fig:1D}a and \ref{fig:1D-2}a), since
\begin{align}
\label{rel-J0}
\lim_{z \to 0} J_0(z)=1\,.
\end{align}
Moreover, the Green function~\eqref{G-BP-1d} 
shows a decreasing oscillation around 
the classical Green function~\eqref{G-w-1d} 
(see Fig.~\ref{fig:1D}b).

\subsection{First-order derivatives of the Green function $G^{L\square}$}
Now, we calculate the first-order time derivative and the gradient 
of the Green function $G^{L\square}$. 
We show that the (first-order) differentiation of the Green function $G^{L\square}$
does not introduce singularities and leads to regular functions. 

\subsubsection{3D} 
The first-order time derivative and the gradient
of the three-dimensional retarded Green function~\eqref{G-BP-3d} read as 
\begin{align}
\label{G-BP-3d-t}
\pd_\tau G_{(3)}^{L\square}(\bm R, \tau)
&=-\frac{c^3 \tau}{4\pi(a_1^2-a_2^2)}\,
\frac{H\big(c\tau-R\big)}{(c^2 \tau^2-R^2)}
\bigg[
J_2 \bigg(\frac{\sqrt{c^2 \tau^2-R^2}}{a_1}\bigg)
-J_2 \bigg(\frac{\sqrt{c^2 \tau^2-R^2}}{a_2}\bigg)
\bigg]\,,\\
\label{G-BP-3d-grad}
\nabla G_{(3)}^{L\square}(\bm R, \tau)
&=\frac{c \bm R}{4\pi(a_1^2-a_2^2)}\,
\frac{H\big(c\tau-R\big)}{(c^2 \tau^2-R^2)}
\bigg[
J_2 \bigg(\frac{\sqrt{c^2 \tau^2-R^2}}{a_1}\bigg)
-J_2 \bigg(\frac{\sqrt{c^2 \tau^2-R^2}}{a_2}\bigg)
\bigg]\,,
\end{align}
where we have used $(J_1(z)/z)'=-J_2(z)/z$. 
$J_2$ is the Bessel function of the first kind of order 2. 
Eqs.~\eqref{G-BP-3d-t} and \eqref{G-BP-3d-grad} consist of
Bessel function terms non-zero inside the light cone.
(see Figs.~\ref{fig:3D}d and \ref{fig:3D-2}d).
The $\delta$-term present in the Bopp-Podolsky theory (first gradient electrodynamics)
vanishes due to the superposition in second gradient electrodynamics.
Taking into account that
\begin{align}
\label{J2-rel}
\lim_{z \to 0} \,\frac{1}{z^2}\,J_2(z)=\frac{1}{8}\,,
\end{align}
it can be seen that the derivatives of the Green function~$G_{(3)}^{L\square}$ 
possess a discontinuity on the light cone 
(see Figs.~\ref{fig:3D}c and \ref{fig:3D-2}c). 
In the neighbourhood of the light cone,
Eqs.~\eqref{G-BP-3d-t} and \eqref{G-BP-3d-grad} 
behave like 
\begin{align}
\label{G-BP-3d-t-LC}
\pd_\tau G_{(3)}^{L\square}(\bm R, \tau)
&\simeq
\frac{c^3 \tau}{32\pi a_1^2a_2^2}\,H\big(c\tau-R\big)\,,\\
\label{G-BP-3d-grad-LC}
\nabla G_{(3)}^{L\square}(\bm R, \tau)
&\simeq
-\frac{c \bm R}{32\pi a_1^2a_2^2}\,H\big(c\tau-R\big)
\,.
\end{align}
Eqs.~\eqref{G-BP-3d-t} and
\eqref{G-BP-3d-grad} show a decreasing oscillation (see Fig.~\ref{fig:3D}d).

\subsubsection{2D} 
The first-order time derivative and the gradient
of the two-dimensional retarded Green function~\eqref{G-BP-2d} read as 
\begin{align}
\label{G-BP-2d-t}
\pd_\tau G_{(2)}^{L\square}(\bm R, \tau)
&=-\frac{c^3\tau}{2\pi}\,
\frac{H\big(c\tau-R\big)}{(c^2\tau^2-R^2)^{\frac{3}{2}}}\, 
\bigg[
1-
\frac{1}{a_1^2-a_2^2}\bigg(
a_1^2  \cos\bigg( \frac{\sqrt{c^2 \tau^2-R^2}}{a_1}\bigg)
-a_2^2  \cos\bigg( \frac{\sqrt{c^2 \tau^2-R^2}}{a_2}\bigg)\nonumber\\
&\qquad
-\sqrt{c^2\tau^2-R^2}\,
\bigg(
a_1  \sin\bigg( \frac{\sqrt{c^2 \tau^2-R^2}}{a_1}\bigg)
-
a_2  \sin\bigg( \frac{\sqrt{c^2 \tau^2-R^2}}{a_2}\bigg)
\bigg)
\bigg)
\bigg] \,,\\
\label{G-BP-2d-grad}
\nabla G_{(2)}^{L\square}(\bm R, \tau)
&=\frac{c\bm R}{2\pi}\,
\frac{H\big(c\tau-R\big)}{(c^2\tau^2-R^2)^{\frac{3}{2}}}\, 
\bigg[
1-
\frac{1}{a_1^2-a_2^2}\bigg(
a_1^2  \cos\bigg( \frac{\sqrt{c^2 \tau^2-R^2}}{a_1}\bigg)
-a_2^2  \cos\bigg( \frac{\sqrt{c^2 \tau^2-R^2}}{a_2}\bigg)\nonumber\\
&\qquad
-\sqrt{c^2\tau^2-R^2}\,
\bigg(
a_1  \sin\bigg( \frac{\sqrt{c^2 \tau^2-R^2}}{a_1}\bigg)
-
a_2  \sin\bigg( \frac{\sqrt{c^2 \tau^2-R^2}}{a_2}\bigg)
\bigg)
\bigg)
\bigg]
\,.
\end{align}
On the light cone, the first derivatives of the Green function~$G_{(2)}^{L\square}$
are zero 
(see Figs.~\ref{fig:2D}c and \ref{fig:2D-2}c) taking into account that
\begin{align}
\lim_{z \to 0} \,\frac{1}{z^3}\,\cos(z)=\frac{1}{z^3}-\frac{1}{2z}
\end{align}
and
\begin{align}
\lim_{z \to 0} \,\frac{1}{z^2}\,\sin(z)=\frac{1}{z}\,.
\end{align}
Eqs.~\eqref{G-BP-2d-t} and \eqref{G-BP-2d-grad}
show a decreasing oscillation around 
the classical singularity (see Fig.~\ref{fig:2D}d).

\subsubsection{1D}

The first-order time derivative and the first-order space derivative
of the one-dimensional retarded Green function~\eqref{G-BP-1d} are given by 
\begin{align}
\label{G-BP-1d-t}
\pd_\tau G_{(1)}^{L\square}(X,\tau)
&=\frac{c^3 \tau}{2(a_1^2-a_2^2)}\, 
\frac{H\big(c\tau-|X|)}{\sqrt{c^2\tau^2-X^2}}\,
\bigg[
a_1 J_1 \bigg( \frac{\sqrt{c^2 \tau^2-X^2}}{a_1}\bigg)
-
a_2 J_1 \bigg( \frac{\sqrt{c^2 \tau^2-X^2}}{a_2}\bigg)
\bigg]
\,,\\
\label{G-BP-1d-grad}
\pd_X G_{(1)}^{L\square}(X,\tau)
&=-\frac{c X}{2(a_1^2-a_2^2)}\, 
\frac{H\big(c\tau-|X|)}{\sqrt{c^2\tau^2-X^2}}\,
\bigg[
a_1 J_1 \bigg( \frac{\sqrt{c^2 \tau^2-X^2}}{a_1}\bigg)
-
a_2 J_1 \bigg( \frac{\sqrt{c^2 \tau^2-X^2}}{a_2}\bigg)
\bigg]
\,,
\end{align}
using $J_0'=-J_1$.
On the light cone, the derivatives of the Green function~$G_{(1)}^{L\square}$,
Eqs.~\eqref{G-BP-1d-t} and \eqref{G-BP-1d-grad}, 
are zero, due to Eq.~\eqref{rel-J1} 
(see Figs.~\ref{fig:1D}c and \ref{fig:1D-2}c).
It can be seen that Eqs.~\eqref{G-BP-1d-t} and \eqref{G-BP-1d-grad}
show a decreasing oscillation (see Fig.~\ref{fig:1D}d) unlike the derivative
of the Green function of the d'Alembert equation given in terms of $\delta\big(c\tau-|X|)$.

\subsection{Green functions for case~2: 
two real and equal length scales}
\label{GF-mul}

In this subsection, we calculate and investigate the Green functions for the
case of real and equal length scale parameters.
For $a_2=a_1$, the bi-Klein-Gordon operator~\eqref{L-op-2} reduces to 
\begin{align}
\label{L-2}
L=[1+a_1\square]^2\,,
\end{align}
which is a double Klein-Gordon operator.

\begin{figure}[t!!]\unitlength1cm
\vspace*{-0.5cm}
\centerline{
\epsfig{figure=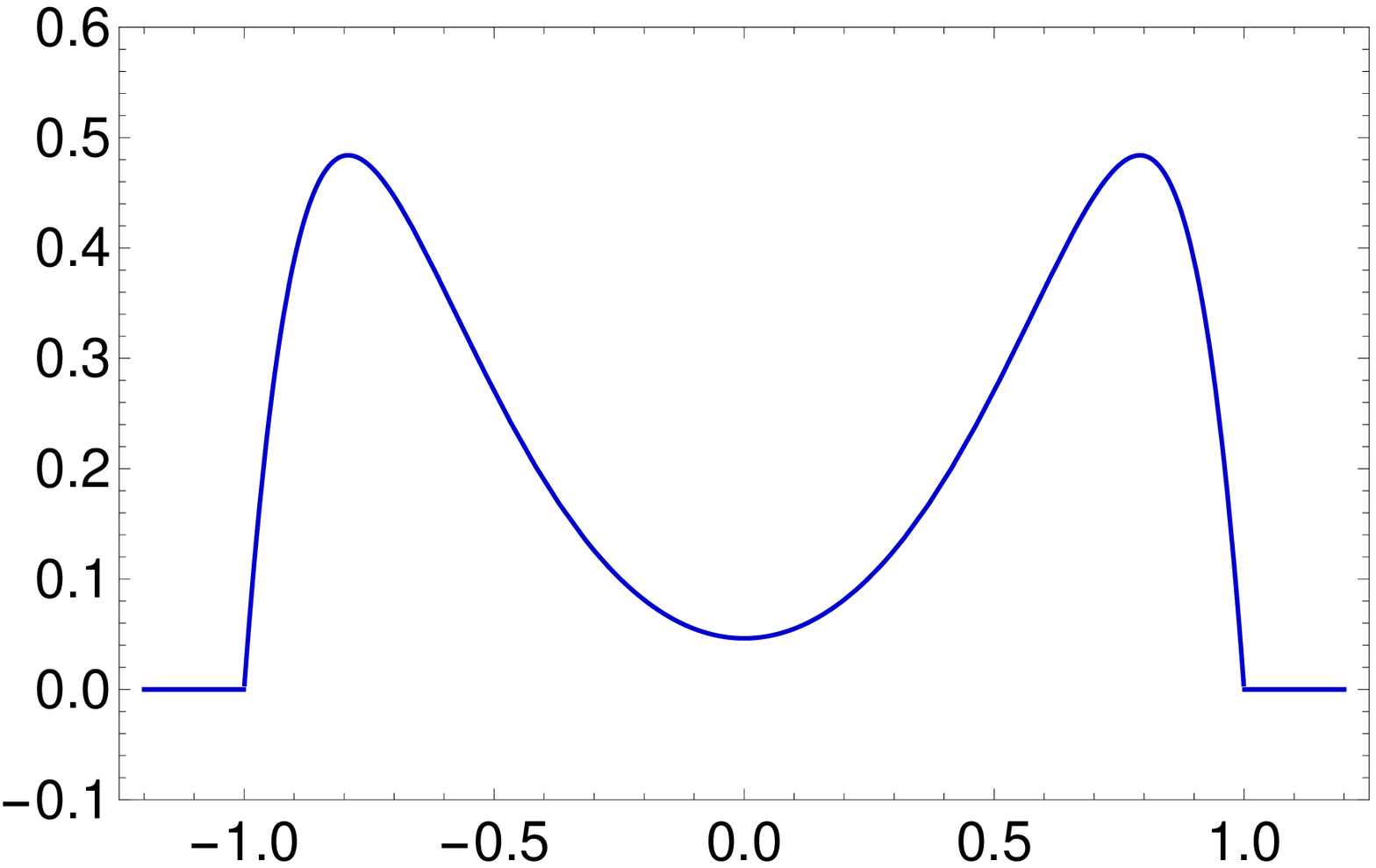,width=7.0cm}
\put(-3.5,-0.4){$X$}
\put(-7.1,-0.3){$\text{(a)}$}
\hspace*{0.4cm}
\put(-0.1,-0.3){$\text{(b)}$}
\put(3.5,-0.4){$\tau$}
\epsfig{figure=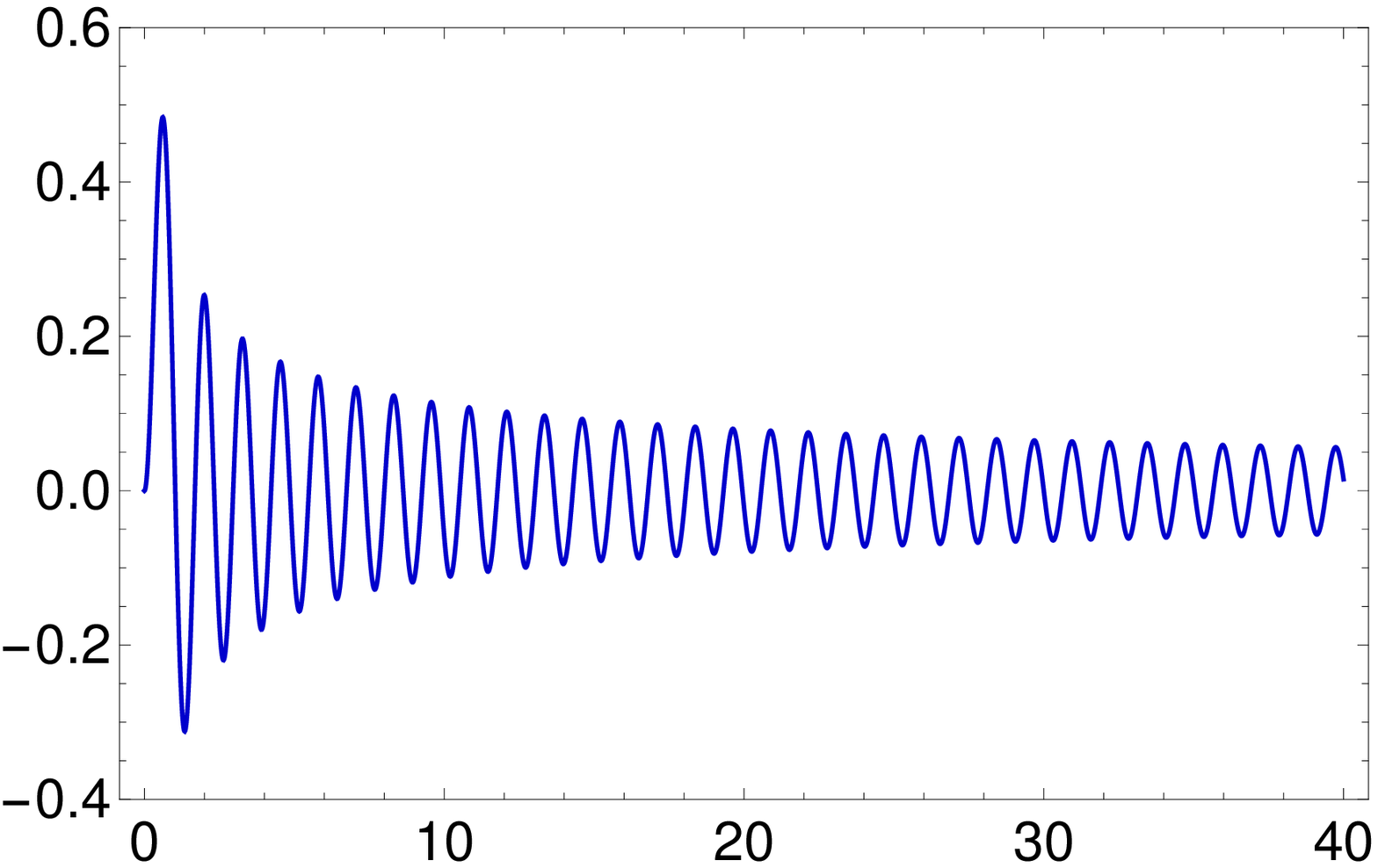,width=7.0cm}
}
\vspace*{0.2cm}
\centerline{
\epsfig{figure=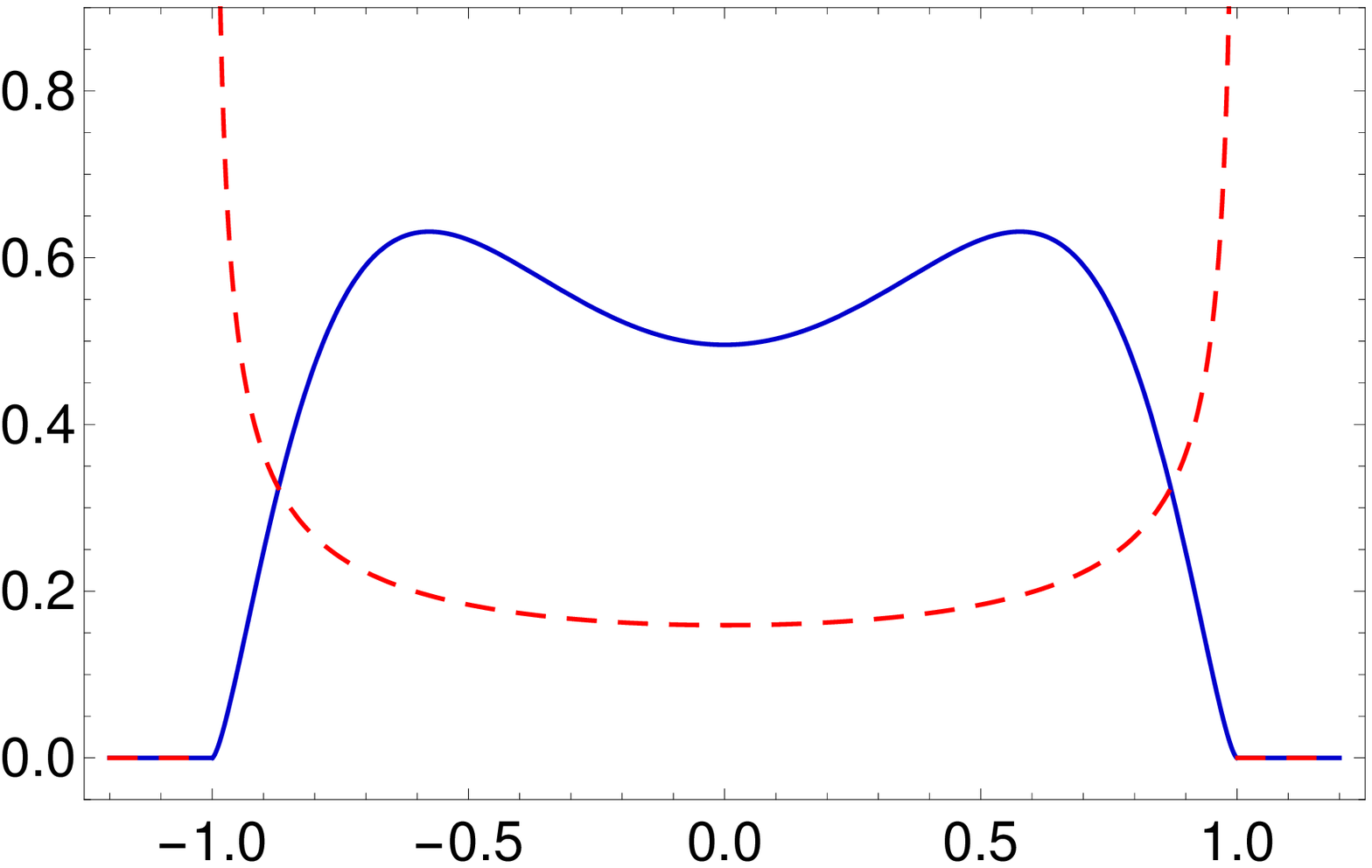,width=6.8cm}
\put(-3.5,-0.4){$X$}
\put(-7.1,-0.3){$\text{(c)}$}
\hspace*{0.4cm}
\put(3.5,-0.4){$\tau$}
\put(-0.1,-0.3){$\text{(d)}$}
\epsfig{figure=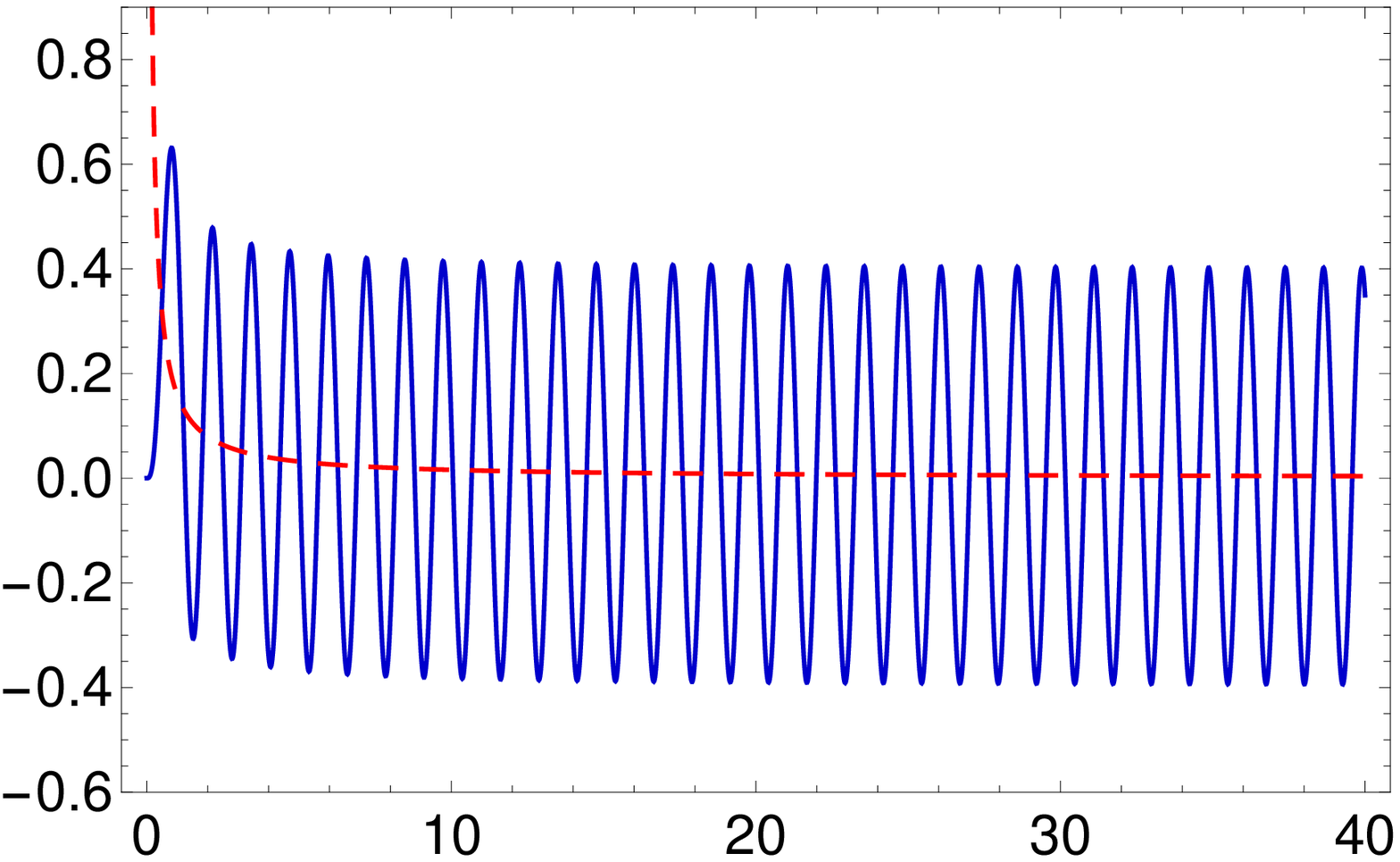,width=7.0cm}
}
\vspace*{0.2cm}
\centerline{
\epsfig{figure=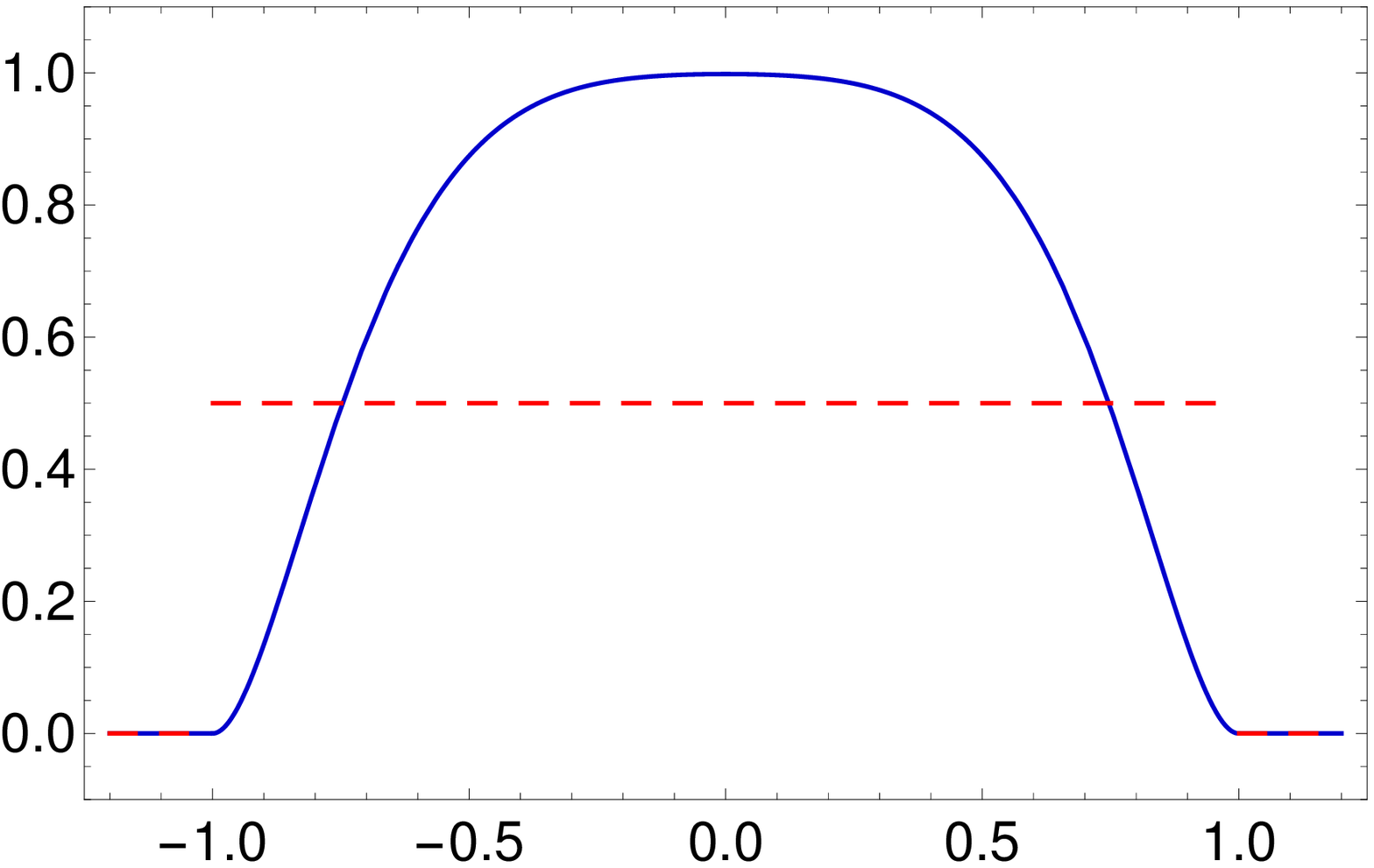,width=6.8cm}
\put(-3.5,-0.4){$X$}
\put(-7.1,-0.3){$\text{(e)}$}
\hspace*{0.4cm}
\put(3.5,-0.4){$\tau$}
\put(-0.1,-0.3){$\text{(f)}$}
\epsfig{figure=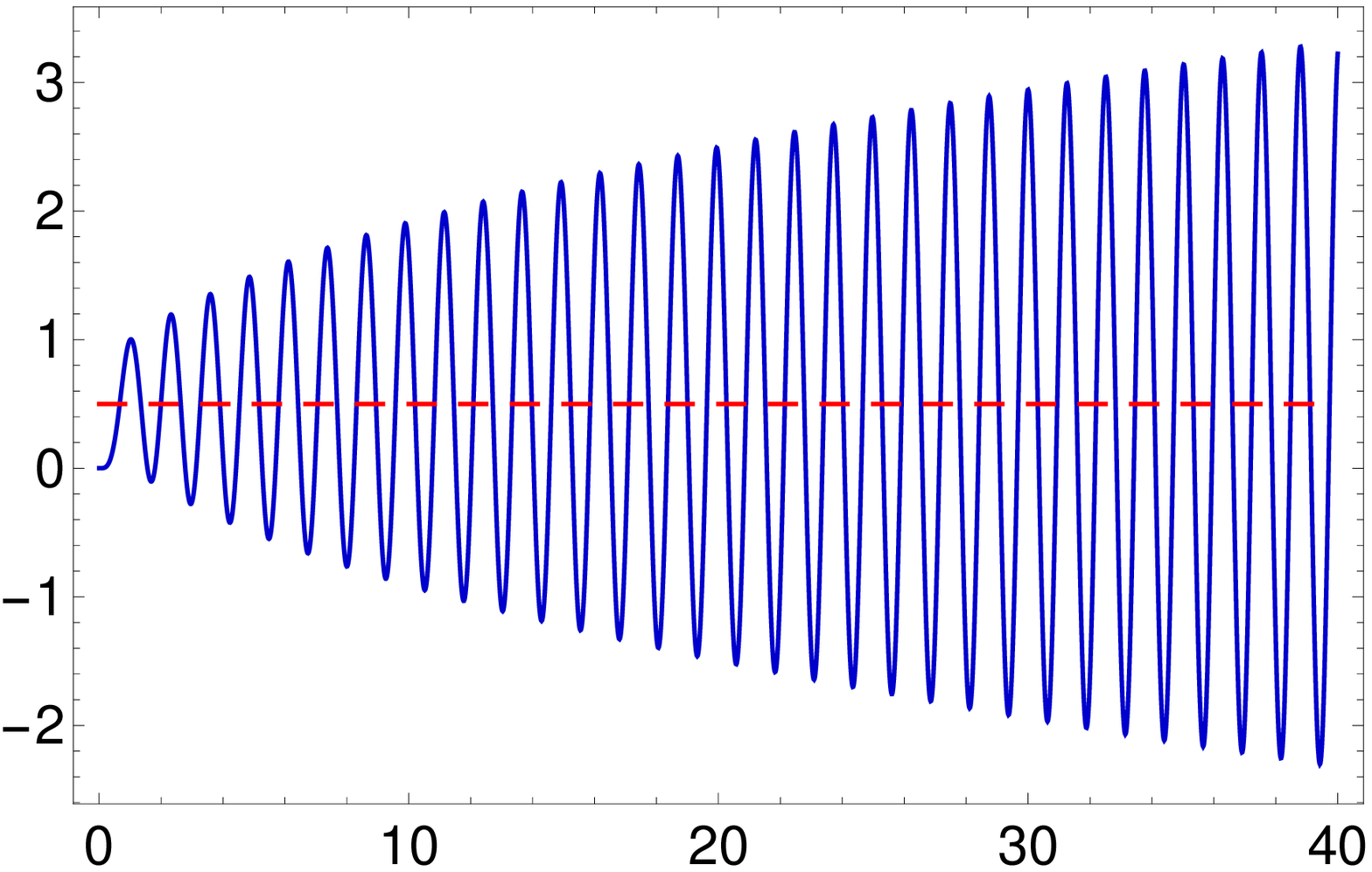,width=7.0cm}
}
\caption{Plots of the Green function for multiple length scale
for $c=1$, $a_1=0.2$: 
(a) $G^{L\square}_{(3)}(\bm R,\tau=1)$ for $Y=Z=0$,
(b) $G^{L\square}_{(3)}(\bm R=0,\tau)$,
(c) $G^{L\square}_{(2)}(\bm R,\tau=1)$ for $Y=0$,
(d) $G^{L\square}_{(2)}(\bm R=0,\tau)$,
(e) $G^{L\square}_{(1)}(X,\tau=1)$,
(f) $G^{L\square}_{(1)}(X=0,\tau)$
(red dashed curves are the classical Green functions).}
\label{fig:GF-mul}
\end{figure}

In the limit $a_2\rightarrow a_1$,
the three-dimensional 
Green functions of the bi-Klein-Gordon operator~\eqref{G-BKG-3d}
and the three-dimensional Green functions of the bi-Klein-Gordon-d'Alembert operator~\eqref{G-BP-3d}
become
\begin{align}
\label{G-BKG-3d-2}
G_{(3)}^{L}(\bm R, \tau)
&=\frac{c}{8\pi a_1^4}\,
H\big(c\tau-R\big)\, 
J_0 \bigg( \frac{\sqrt{c^2 \tau^2-R^2}}{a_1}\bigg)\,,\\
\label{G-BP-3d-2}
G_{(3)}^{L\square}(\bm R, \tau)
&=\frac{c}{8\pi a_1^2}\,
H\big(c\tau-R\big)\,
J_2 \bigg( \frac{\sqrt{c^2 \tau^2-R^2}}{a_1}\bigg)\,.
\end{align}
The Bessel functions $J_0$ and $J_2$ in Eqs.~\eqref{G-BKG-3d-2}
and \eqref{G-BP-3d-2}, respectively, give a slowly decreasing oscillation
of the Green functions (see Fig.~\ref{fig:GF-mul}b).

In the limit $a_2\rightarrow a_1$,
the two-dimensional 
Green functions of the bi-Klein-Gordon operator~\eqref{G-BKG-2d}
and the two-dimensional Green functions of the bi-Klein-Gordon-d'Alembert operator~\eqref{G-BP-2d}
reduce to 
\begin{align}
\label{G-BKG-2d-2}
G_{(2)}^{L}(\bm R, \tau)
&=\frac{c}{4\pi a_1^3}\,
H\big(c\tau-R\big)\,
\sin\bigg( \frac{\sqrt{c^2 \tau^2-R^2}}{a_1}\bigg)
\,,\\
\label{G-BP-2d-2}
G_{(2)}^{L\square}(\bm R, \tau)
&=\frac{c}{2\pi}\,
\frac{H\big(c\tau-R\big)}{\sqrt{c^2\tau^2-R^2}}\, 
\bigg[1-
\cos\bigg( \frac{\sqrt{c^2 \tau^2-R^2}}{a_1}\bigg)
-\frac{\sqrt{c^2 \tau^2-R^2}}{2 a_1}\,
\sin\bigg( \frac{\sqrt{c^2 \tau^2-R^2}}{a_1}\bigg)
\bigg] \,.
\end{align}
The $\sin$-term in Eqs.~\eqref{G-BKG-2d-2}
and \eqref{G-BP-2d-2} gives rise to an oscillation
of the Green functions (see Fig.~\ref{fig:GF-mul}d).

In the limit $a_2\rightarrow a_1$,
the one-dimensional 
Green functions of the bi-Klein-Gordon operator~\eqref{G-BKG-1d}
and the one-dimensional Green functions of the bi-Klein-Gordon-d'Alembert operator~\eqref{G-BP-1d}
become
\begin{align}
\label{G-BKG-1d-2}
G_{(1)}^{L}(X,\tau)
&=\frac{c}{4 a_1^3}\,
H\big(c\tau-|X|\big)\, \sqrt{c^2 \tau^2-X^2}\,
J_1 \bigg( \frac{\sqrt{c^2 \tau^2-X^2}}{a_1}\bigg)
\,,\\
\label{G-BP-1d-2}
G_{(1)}^{L\square}(X,\tau)
&=\frac{c}{2}\, H\big(c\tau-|X|)
\bigg[1-
J_0 \bigg( \frac{\sqrt{c^2 \tau^2-X^2}}{a_1}\bigg)
-\frac{\sqrt{c^2 \tau^2-X^2}}{2 a_1}\,
J_1 \bigg( \frac{\sqrt{c^2 \tau^2-X^2}}{a_1}\bigg)
\bigg]\,.
\end{align}
The term $z J_1(z)$ in Eqs.~\eqref{G-BKG-1d-2}
and \eqref{G-BP-1d-2} causes a slowly increasing oscillation
of the Green functions (see Fig.~\ref{fig:GF-mul}f).

Eqs.~\eqref{G-BKG-3d-2}, \eqref{G-BKG-2d-2} and \eqref{G-BKG-1d-2} are in
agreement with the corresponding expressions of the Green function of the
iterated Klein-Gordon equation given in~\citep{Schwartz,Jager,Kanwal}.

\section{Retarded potentials and retarded electromagnetic field strengths}
\label{sec4}
In this section, we derive the retarded potentials and the retarded electromagnetic field strengths in the framework of 
second gradient electrodynamics.
The solutions of the field equations, which are  
based on the retarded Green functions, lead to retarded fields 
(retarded potentials and retarded electromagnetic field strengths)
in the form of retarded integrals.
Those retarded integrals reflect the 
phenomenon of the ``finite signal speed" for the propagation of 
the electromagnetic fields (e.g.~\citep{Jefimenko}).

\subsection{Retarded potentials}

In second gradient electrodynamics,
the retarded electromagnetic potentials are the solutions of the inhomogeneous partial differential equations of sixth order~\eqref{phi-w} and \eqref{A-w}.
For zero initial conditions, they are given as convolution of the
retarded Green function $G^{L\square}$ and 
given charge and current densities ($\rho$, $\bm J$) 
\begin{align}
\label{phi-rp}
\phi&=\frac{1}{\varepsilon_0}\, G^{L\square}*\rho\,,\\
\label{A-rp}
\bm A&=\mu_0\, G^{L\square}*\bm J\,.
\end{align}
Explicitly, the convolution integrals~\eqref{phi-rp} and \eqref{A-rp} are given by 
\begin{align}
\label{phi-rp-2}
\phi_{(n)}(\bm r,t)&=\frac{1}{\varepsilon_0}\, \int_{-\infty}^t\d t'\int_{\Bbb R^n} \d \bm r'\, G_{(n)}^{L\square}(\bm r-\bm r',t-t')\,\rho(\bm r',t')\,,\\
\label{A-rp-2}
\bm A_{(n)}(\bm r,t)&=\mu_0 \int_{-\infty}^t\d t'\int_{\Bbb R^n} \d \bm r'\, G_{(n)}^{L\square}(\bm r-\bm r',t-t') \,\bm J(\bm r',t')\,,
\end{align}
where $\bm r'$ is the source point and $\bm r$ is the field point. Here $n$ denotes the spatial dimension.

\subsubsection{3D} 

If we substitute the three-dimensional Green function~\eqref{G-BP-3d} into Eqs.~\eqref{phi-rp-2} and \eqref{A-rp-2}, 
then the three-dimensional retarded electromagnetic potentials become
\begin{align}
\label{phi-ret-3d}
\phi_{(3)}(\bm r,t)
&=\frac{c}{4\pi\varepsilon_0(a_1^2-a_2^2)}\int_{-\infty}^{t-R/c}\d t'\int_{\Bbb R^3} \d \bm r'\,
\frac{\rho(\bm r',t')}{\sqrt{c^2 \tau^2-R^2}}\,
\bigg[
a_1 J_1 \bigg( \frac{\sqrt{c^2 \tau^2-R^2}}{a_1}\bigg)
\nonumber\\
&\hspace*{8cm}
-a_2 J_1 \bigg( \frac{\sqrt{c^2 \tau^2-R^2}}{a_2}\bigg)
\bigg]
\end{align}
and
\begin{align}
\label{A-ret-3d}
\bm A_{(3)}(\bm r,t)
&=\frac{\mu_0 c}{4\pi(a_1^2-a_2^2)}
\int_{-\infty}^{t-R/c}\d t'\int_{\Bbb R^3} \d \bm r'\,
\frac{\bm J(\bm r',t')}{\sqrt{c^2 \tau^2-R^2}}\, 
\bigg[
a_1 J_1 \bigg( \frac{\sqrt{c^2 \tau^2-R^2}}{a_1}\bigg)
\nonumber\\
&\hspace*{8cm}
-a_2 J_1 \bigg( \frac{\sqrt{c^2 \tau^2-R^2}}{a_2}\bigg)
\bigg]
\end{align}
since $H(c\tau -R)=0$ for $t'>t-R/c$.
In second gradient electrodynamics,  the three-dimensional 
retarded potentials~\eqref{phi-ret-3d}
and \eqref{A-ret-3d} possess an afterglow,
since they draw contribution emitted at all times $t'$ from $-\infty$ up to $t-R/c$.

\subsubsection{2D} 

Inserting the two-dimensional Green function~\eqref{G-BP-2d} into Eqs.~\eqref{phi-rp-2} and \eqref{A-rp-2}, 
the two-dimensional retarded electromagnetic potentials read as 
\begin{align}
\label{phi-ret-2d}
\phi_{(2)}(\bm r,t)
&=\frac{c}{2\pi\varepsilon_0}\int_{-\infty}^{t-R/c}\d t'\int_{\Bbb R^2} \d \bm r'\,
\frac{\rho(\bm r',t')}{\sqrt{c^2\tau^2-R^2}}\, 
\bigg[1-
\frac{1}{a_1^2-a_2^2}\bigg(
a_1^2 \cos\bigg( \frac{\sqrt{c^2 \tau^2-R^2}}{a_1}\bigg)
\nonumber\\
&\hspace*{8cm}
-a_2^2 \cos\bigg( \frac{\sqrt{c^2 \tau^2-R^2}}{a_2}\bigg)\bigg)
\bigg] 
\end{align}
and
\begin{align}
\label{A-ret-2d}
\bm A_{(2)}(\bm r,t)
&=\frac{\mu_0 c}{2\pi}\int_{-\infty}^{t-R/c}\d t'\int_{\Bbb R^2} \d \bm r'\,
\frac{\bm J(\bm r',t')}{\sqrt{c^2\tau^2-R^2}}\, 
\bigg[1-
\frac{1}{a_1^2-a_2^2}\bigg(
a_1^2\cos\bigg(\frac{\sqrt{c^2 \tau^2-R^2}}{a_1}\bigg)
\nonumber\\
&\hspace*{8cm}
-a_2^2\cos\bigg(\frac{\sqrt{c^2 \tau^2-R^2}}{a_2}\bigg)\bigg)
\bigg] 
\end{align}
since $H(c\tau -R)=0$ for $t'>t-R/c$.
It can be seen that the two-dimensional 
retarded potentials~\eqref{phi-ret-2d}
and \eqref{A-ret-2d} possess an afterglow,
since they draw contribution emitted at all times $t'$ from $-\infty$ up to $t-R/c$.

\subsubsection{1D}

In one-dimensional electrodynamics,  
the potentials $\phi_{(1)}$ and $A_{(1)}$, and the current
density $J$ are scalar fields.

Now inserting the one-dimensional Green function~\eqref{G-BP-1d} into Eqs.~\eqref{phi-rp-2} and \eqref{A-rp-2}, 
the one-dimensional retarded electromagnetic potentials read as 
\begin{align}
\label{phi-ret-1d}
\phi_{(1)}(x,t)
&=\frac{c}{2\varepsilon_0}\int_{-\infty}^{t-|X|/c} \d t'\int_{-\infty}^\infty\d x'
\,\rho(x',t')
\bigg[1-
\frac{1}{a_1^2-a_2^2}\bigg(
a_1^2 J_0 \bigg( \frac{\sqrt{c^2\tau^2-X^2}}{a_1}\bigg)
\nonumber\\
&\hspace*{8cm}
-a_2^2 J_0 \bigg( \frac{\sqrt{c^2\tau^2-X^2}}{a_2}\bigg)\bigg)
\bigg]
\end{align}
and
\begin{align}
\label{A-ret-1d}
A_{(1)}(x,t)
&=\frac{\mu_0 c}{2}\int_{-\infty}^{t-|X|/c}\d t'\int_{-\infty}^\infty\d x'
\, J(x',t')
\bigg[1-
\frac{1}{a_1^2-a_2^2}\bigg(
a_1^2 J_0 \bigg( \frac{\sqrt{c^2\tau^2-X^2}}{a_1}\bigg)
\nonumber\\
&\hspace*{8cm}
-a_2^2 J_0 \bigg( \frac{\sqrt{c^2\tau^2-X^2}}{a_2}\bigg)\bigg)
\bigg]
\end{align}
since $H(c\tau -|X|)=0$ for $t'>t-|X|/c$.
The one-dimensional 
retarded potentials~\eqref{phi-ret-1d}
and \eqref{A-ret-1d} draw contribution emitted at all times $t'$ from $-\infty$ up to $t-|X|/c$.

Therefore, in second gradient electrodynamics,  the retarded potentials possess an afterglow in 1D, 2D and 3D since they draw contribution emitted
at all times $t'$ from $-\infty$ up to $t-R/c$ 
unlike in the classical Maxwell electrodynamics where only 
the retarded potentials possess an afterglow in 1D and 2D (see, e.g., \citep{Barton,Wl}).

\subsection{Retarded electromagnetic field strengths}

Inserting Eqs.~\eqref{phi-rp} and \eqref{A-rp} into 
the electromagnetic fields~\eqref{E} and \eqref{B} or solving 
Eqs.~\eqref{E-w} and \eqref{B-w}, the electromagnetic fields ($\bm E$, $\bm B$)
are given by the convolution of the Green function~$G^{L\square}$ and 
the given charge and current densities ($\rho$, $\bm J$) 
\begin{align}
\label{E-rp1}
\bm E&=-\frac{1}{\varepsilon_0}\,\bigg(\nabla G^{L\square}*\rho+\frac{1}{c^2}\,
\pd_t G^{L\square}*\bm J\bigg)\,,\\
\label{B-rp1}
\bm B&=\mu_0\, \nabla\times \big( G^{L\square}*\bm J\big)\,.
\end{align}
Explicitly, the convolution integrals~\eqref{E-rp1} and \eqref{B-rp1} read as 
\begin{align}
\label{E-rp}
\bm E_{(n)}(\bm r,t)&=-\frac{1}{\varepsilon_0}\, \int_{-\infty}^t\d t'\int_{\Bbb R^n}
\d \bm r'\, 
\Big(\nabla G_{(n)}^{L\square}(\bm r-\bm r',t-t')\,\rho(\bm r',t')
\nonumber\\
&\qquad\qquad\qquad\qquad
+\frac{1}{c^2} \pd_t  G_{(n)}^{L\square}(\bm r-\bm r',t-t')\, \bm J(\bm r',t')\Big)
\,,\\
\label{B-rp}
\bm B_{(n)}(\bm r,t)&=\mu_0 \int_{-\infty}^t\d t'\int_{\Bbb R^n} \d \bm r'\,
\nabla G_{(n)}^{L\square}(\bm r-\bm r',t-t') \times \bm J(\bm r',t')\,.
\end{align}

\subsubsection{3D}

Substituting the derivatives of the  three-dimensional 
Green function~\eqref{G-BP-3d-t} 
and \eqref{G-BP-3d-grad} into Eqs.~\eqref{E-rp} and \eqref{B-rp}, 
the three-dimensional retarded electromagnetic field strengths read as 
\begin{align}
\label{E-ret-3d}
\bm E_{(3)}(\bm r,t)
&=-\frac{c}{4\pi\varepsilon_0 (a_1^2-a_2^2)}\,
\int_{-\infty}^{t-R/c}\d t'\int_{\Bbb R^3}\d \bm r'\, 
\frac{\big[\bm R \rho(\bm r',t')-\tau \bm J(\bm r',t')\big]}
{(c^2 \tau^2-R^2)}\, 
\bigg[
J_2 \bigg(\frac{\sqrt{c^2 \tau^2-R^2}}{a_1}\bigg)
\nonumber\\
&\hspace*{9cm}
-J_2 \bigg(\frac{\sqrt{c^2 \tau^2-R^2}}{a_2}\bigg)
\bigg]
\end{align}
and 
\begin{align}
\label{B-ret-3d}
\bm B_{(3)}(\bm r,t)
&=\frac{\mu_0 c}{4\pi(a_1^2-a_2^2)}\,
\int_{-\infty}^{t-R/c}\d t'\int_{\Bbb R^3}\d \bm r'\, 
\frac{\bm R \times \bm J(\bm r',t')}{(c^2 \tau^2-R^2)}\, 
\bigg[
J_2 \bigg(\frac{\sqrt{c^2 \tau^2-R^2}}{a_1}\bigg)
\nonumber\\
&\hspace*{8cm}
-J_2 \bigg(\frac{\sqrt{c^2 \tau^2-R^2}}{a_2}\bigg)
\bigg]
\,.
\end{align}
The three-dimensional retarded electromagnetic field strengths~\eqref{E-ret-3d} and \eqref{B-ret-3d}
draw contribution emitted at all times $t'$ from $-\infty$ up to $t-R/c$.

\subsubsection{2D}

In two-dimensional electrodynamics, 
the electric field strength $E_{(2)}=(E_x,E_y)$ 
is a two-dimensional vector field and 
the magnetic field strength is a scalar field 
$B_{(2)}=\nabla\times\bm A_{(2)}=\epsilon_{ij}\pd_i A_j$ where
$\epsilon_{ij}$ denotes the two-dimensional Levi-Civita tensor
(see also~\citep{HO}).

If we substitute the derivatives of the Green function~\eqref{G-BP-2d-t} 
and \eqref{G-BP-2d-grad} into Eqs.~\eqref{E-rp} and \eqref{B-rp}, 
the two-dimensional retarded electromagnetic field strengths read as 
\begin{align}
\label{E-ret-2d}
\bm E_{(2)}(\bm r,t)
&=-\frac{c}{2\pi \varepsilon_0}\int_{-\infty}^{t-R/c}\d t'\int_{\Bbb R^2}\d \bm r'\, 
\frac{\big[\bm R \rho(\bm r',t')-\tau \bm J(\bm r',t')\big]}{(c^2\tau^2-R^2)^\frac{3}{2}}\, 
\bigg[1
\nonumber\\
&\qquad
-\frac{1}{a_1^2-a_2^2}\bigg(
a_1^2 \cos\bigg( \frac{\sqrt{c^2 \tau^2-R^2}}{a_1}\bigg)
-a_2^2 \cos\bigg( \frac{\sqrt{c^2 \tau^2-R^2}}{a_2}\bigg)
\nonumber\\
&\qquad
-\sqrt{c^2\tau^2-R^2}\,
\bigg(
a_1 \sin\bigg( \frac{\sqrt{c^2 \tau^2-R^2}}{a_1}\bigg)
-
a_2 \sin\bigg( \frac{\sqrt{c^2 \tau^2-R^2}}{a_2}\bigg)
\bigg)\bigg)
\bigg]
\end{align}
and
\begin{align}
\label{B-ret-2d}
&B_{(2)}(\bm r,t)
=\frac{\mu_0 c}{2\pi}\int_{-\infty}^{t-R/c}\d t'\int_{\Bbb R^2}\d \bm r'\, 
\frac{\bm R\times \bm J(\bm r',t')}{(c^2\tau^2-R^2)^\frac{3}{2}}\, 
\bigg[1
-\frac{1}{a_1^2-a_2^2}\bigg(
a_1^2 \cos\bigg( \frac{\sqrt{c^2 \tau^2-R^2}}{a_1}\bigg)
\nonumber\\
&\quad
-a_2^2 \cos\bigg( \frac{\sqrt{c^2 \tau^2-R^2}}{a_2}\bigg)
-\sqrt{c^2\tau^2-R^2}\,
\bigg(
a_1 \sin\bigg( \frac{\sqrt{c^2 \tau^2-R^2}}{a_1}\bigg)
-
a_2 \sin\bigg( \frac{\sqrt{c^2 \tau^2-R^2}}{a_2}\bigg)
\bigg)\bigg)
\bigg] \,,
\end{align}
where $\bm R\times \bm J=\epsilon_{ij} R_i J_j$. 
In the second gradient electrodynamics,  the two-dimensional 
retarded electromagnetic field strengths~\eqref{E-ret-2d}
and \eqref{B-ret-2d} possess an afterglow,
since they draw contribution emitted at all times $t'$ from $-\infty$ up to $t-R/c$.

\subsubsection{1D}

This version of one-dimensional electrodynamics has a scalar electric
field and no magnetic field (see, e.g.,~\cite{Galic} for classical
electrodynamics in one spatial dimension). 

Substituting the derivatives of the Green function~\eqref{G-BP-1d-t} 
and \eqref{G-BP-1d-grad} into Eqs.~\eqref{E-rp} and \eqref{B-rp}, 
the one-dimensional retarded electromagnetic field strengths read as 
\begin{align}
\label{E-ret-1d}
E_{(1)}(x,t)
&=\frac{c}{2\varepsilon_0(a_1^2-a_2^2)}\int_{-\infty}^{t-|X|/c}
\d t'\int_{-\infty}^\infty\d x'\, 
\frac{\big[X\rho(x',t')-\tau J(x',t')\big]}{\sqrt{c^2\tau^2-X^2}}\,
\bigg[
a_1 J_1 \bigg( \frac{\sqrt{c^2\tau^2-X^2}}{a_1}\bigg)
\nonumber\\
&\hspace*{8.5cm}
-a_2 J_1 \bigg( \frac{\sqrt{c^2\tau^2-X^2}}{a_2}\bigg)
\bigg]\,,
\\
\label{B-ret-1d}
B_{(1)}(x,t)
&=0\,.
\end{align}
Thus, the one-dimensional 
retarded electric field strength~\eqref{E-ret-1d}
possesses an afterglow,
since it draws contribution emitted at all times $t'$ from $-\infty$ up to $t-|X|/c$.

\section{Generalized Li{\'e}nard-Wiechert fields:
electromagnetic fields of a non-uniformly moving point charge}
\label{sec5}

In this section, 
we consider in the framework of second gradient electrodynamics a non-uniformly moving point charge carrying the charge $q$ 
at the position $\Bs(t)$ 
with velocity $\BV(t)=\pd_t{\Bs}(t)=\dot{\Bs}(t)$
which is less than the speed of light: $|\BV|<c$.
The electric charge density and the electric current density vector of a charged point particle are given by
\begin{align}
\label{J}
\rho(\rr,t)&=q \,\delta(\rr-\Bs(t))\,,\qquad
\BJ(\rr,t)=q \BV(t)\, \delta(\rr-\Bs(t))
=\rho(\rr,t) \BV(t)\,.
\end{align}

\subsection{Generalized Li{\'e}nard-Wiechert potentials}

\subsubsection{3D} 

The three-dimensional generalized Li{\'e}nard-Wiechert potentials of a charged point particle are obtained 
by substituting Eq.~(\ref{J}) into Eqs.~(\ref{phi-ret-3d}) and (\ref{A-ret-3d}) 
and performing the spatial integration
\begin{align}
\label{phi-LW-3d}
\phi_{(3)}(\bm r,t)
&=\frac{q c}{4\pi\varepsilon_0(a_1^2-a_2^2)}\int_{-\infty}^\tR\d t'\,
\frac{1}{\sqrt{c^2 \tau^2-R^2(t')}}\, 
\bigg[
a_1 J_1 \bigg(\frac{\sqrt{c^2 \tau^2-R^2(t')}}{a_1}\bigg) 
\nonumber\\
&\hspace*{7cm}
-a_2 J_1 \bigg(\frac{\sqrt{c^2 \tau^2-R^2(t')}}{a_2}\bigg)
\bigg] 
\end{align}
and 
\begin{align}
\label{A-LW-3d}
\bm A_{(3)}(\bm r,t)
&=\frac{\mu_0 qc}{4\pi(a_1^2-a_2^2)}\int_{-\infty}^\tR\d t'\,
\frac{\bm V(t')}{\sqrt{c^2 \tau^2-R^2(t')}}\, 
\bigg[
a_1 J_1 \bigg(\frac{\sqrt{c^2 \tau^2-R^2(t')}}{a_1}\bigg) 
\nonumber\\
&\hspace*{7cm}
-a_2 J_1 \bigg(\frac{\sqrt{c^2 \tau^2-R^2(t')}}{a_2}\bigg)
\bigg] \,,
\end{align}
where $\bm R(t')=\bm r-\bm s(t')$
and the retarded time~$\tR$ which is the root of the equation
\begin{align}
\label{tR-3d}
\big[x-s_x(\tR)\big]^2+\big[y-s_y(\tR)\big]^2+\big[z-s_z(\tR)\big]^2
-c^2(t-\tR)^2=0\,.
\end{align}
For $|\BV|<c$, there is only one solution of Eq.~\eqref{tR-3d} which is the retarded time $\tR$.

\subsubsection{2D} 

If we insert Eq.~(\ref{J}) into Eqs.~(\ref{phi-ret-2d}) and (\ref{A-ret-2d}) 
and perform the spatial integration,
the two-dimensional generalized Li{\'e}nard-Wiechert potentials are obtained
as 
\begin{align}
\label{phi-LW-2d}
\phi_{(2)}(\bm r,t)
&=\frac{qc}{2\pi\varepsilon_0}\int_{-\infty}^\tR\d t'\,
\frac{1}{\sqrt{c^2\tau^2-R^2(t')}}\, 
\bigg[1-
\frac{1}{a_1^2-a_2^2}\,
\bigg(
a_1^2 \cos\bigg( \frac{\sqrt{c^2 \tau^2-R^2(t')}}{a_1}\bigg)
\nonumber\\
&\hspace*{8cm}
-a_2^2 \cos\bigg( \frac{\sqrt{c^2 \tau^2-R^2(t')}}{a_2}\bigg)
\bigg)
\bigg] 
\end{align}
and
\begin{align}
\label{A-LW-2d}
\bm A_{(2)}(\bm r,t)
&=\frac{\mu_0 qc}{2\pi}\int_{-\infty}^\tR\d t'\,
\frac{\bm V(t')}{\sqrt{c^2\tau^2-R^2(t')}}\, 
\bigg[1-
\frac{1}{a_1^2-a_2^2}\,
\bigg(
a_1^2 \cos\bigg( \frac{\sqrt{c^2 \tau^2-R^2(t')}}{a_1}\bigg)
\nonumber\\
&\hspace*{8cm}
-a_2^2 \cos\bigg( \frac{\sqrt{c^2 \tau^2-R^2(t')}}{a_2}\bigg)
\bigg)
\bigg]\,,
\end{align}
where $\bm R(t')=\bm r-\bm s(t')$
and the retarded time~$\tR$ is the root of the equation
\begin{align}
\label{tR-2d}
\big[x-s_x(\tR)\big]^2+\big[y-s_y(\tR)\big]^2-c^2(t-\tR)^2=0\,.
\end{align}

\subsubsection{1D} 

The one-dimensional generalized Li{\'e}nard-Wiechert potentials are obtained by
inserting Eq.~(\ref{J}) into Eqs.~(\ref{phi-ret-1d}) and (\ref{A-ret-1d}) and performing 
the spatial integration
\begin{align}
\label{phi-LW-1d}
\phi_{(1)}(x,t)
&=\frac{qc}{2\varepsilon_0}\int_{-\infty}^\tR\d t'\, 
\bigg[1-
\frac{1}{a_1^2-a_2^2}\,\bigg(a_1^2 J_0 \bigg( \frac{\sqrt{c^2\tau^2-X^2(t')}}{a_1}\bigg)
-a_2^2 J_0 \bigg( \frac{\sqrt{c^2\tau^2-X^2(t')}}{a_2}\bigg)
\bigg)
\bigg]
\end{align}
and 
\begin{align}
\label{A-LW-1d}
A_{(1)}(x,t)
&=\frac{\mu_0 qc}{2}\int_{-\infty}^\tR\d t'\,   V(t')
\bigg[1-
\frac{1}{a_1^2-a_2^2}\,\bigg(a_1^2 J_0 \bigg( \frac{\sqrt{c^2\tau^2-X^2(t')}}{a_1}\bigg)
-a_2^2 J_0 \bigg( \frac{\sqrt{c^2\tau^2-X^2(t')}}{a_2}\bigg)
\bigg)
\bigg]\,,
\end{align}
where $X(t')=x-s(t')$
and $\tR$ is the retarded time being the root of the equation
\begin{align}
\label{tR-1d}
\big[x-s(\tR)\big]^2-c^2(t-\tR)^2=0\,.
\end{align}

Such as the retarded potentials in second gradient electrodynamics, 
the generalized  Li{\'e}nard-Wiechert potentials in 3D, 2D and 1D,
Eqs.~(\ref{phi-LW-3d}), (\ref{A-LW-3d}), (\ref{phi-LW-2d}), (\ref{A-LW-2d})
and (\ref{phi-LW-1d}), (\ref{A-LW-1d})  depend on the entire history of the
point charge up to the retarded time $\tR$ and contain ``tail terms".

\subsection{Generalized Li{\'e}nard-Wiechert form of the electromagnetic field strengths}

\subsubsection{3D}

The three-dimensional electromagnetic fields of a charged point particle in 
generalized Li{\'e}nard-Wiechert form are obtained by
substituting Eq.~(\ref{J}) into Eqs.~(\ref{E-ret-3d}) and (\ref{B-ret-3d}) 
and performing the spatial integration
\begin{align}
\label{E-LW-3d}
\bm E_{(3)}(\bm r,t)
&=-\frac{qc}{4\pi\varepsilon_0 (a_1^2-a_2^2)}\,
\int_{-\infty}^\tR\d t'\,
\frac{\bm R(t')-\tau \bm V(t')}{(c^2 \tau^2-R^2(t'))}\, 
\bigg[
J_2 \bigg(\frac{\sqrt{c^2 \tau^2-R^2(t')}}{a_1}\bigg)
\nonumber\\
&\hspace*{7.5cm}
-J_2 \bigg(\frac{\sqrt{c^2 \tau^2-R^2(t')}}{a_2}\bigg)
\bigg]
\end{align}
and
\begin{align}
\label{B-LW-3d}
\bm B_{(3)}(\bm r,t)
&=\frac{\mu_0 qc}{4\pi(a_1^2-a_2^2)}\,
\int_{-\infty}^{\tR} 
\d t'\, 
\frac{\bm R(t') \times \bm V(t') }{(c^2 \tau^2-R^2(t'))}\, 
\bigg[
J_2 \bigg(\frac{\sqrt{c^2 \tau^2-R^2(t')}}{a_1}\bigg)
\nonumber\\
&\hspace*{7cm}
-J_2 \bigg(\frac{\sqrt{c^2 \tau^2-R^2(t')}}{a_2}\bigg)
\bigg]\,.
\end{align}
Eqs.~\eqref{E-LW-3d} and \eqref{B-LW-3d}
draw contribution emitted at all times $t'$ from $-\infty$ up to the retarded time $\tR$.
Unlike in the Bopp-Podolsky theory (see, e.g., \citep{Perlick2015,Lazar19}),
no directional discontinuity is present in the electromagnetic field strengths~\eqref{E-LW-3d} and \eqref{B-LW-3d}.
No ad-hoc averaging procedure of the electromagnetic field strength 
as done by~\citet{Perlick2015} in the Bopp-Podolsky electrodynamics 
is needed in second gradient electrodynamics. 
Therefore, the three-dimensional electromagnetic field strengths, Eqs.~\eqref{E-LW-3d} and \eqref{B-LW-3d}, 
are singularity-free unlike the three-dimensional electromagnetic field strengths in the Bopp-Podolsky 
electrodynamics (see also~\citep{Lazar19}).

\subsubsection{2D}

The two-dimensional electromagnetic fields in the 
generalized Li{\'e}nard-Wiechert form are obtained by 
substituting Eq.~(\ref{J}) into Eqs.~(\ref{E-ret-2d}) and (\ref{B-ret-2d}) 
and performing the spatial integration
\begin{align}
\label{E-LW-2d}
&\bm E_{(2)}(\bm r,t)
=-\frac{qc}{2\pi \varepsilon_0}\int_{-\infty}^\tR\d t'\,
\frac{\bm R(t')-\tau \bm V(t')}{(c^2\tau^2-R^2(t'))^\frac{3}{2}}\, 
\bigg[
1-\frac{1}{a_1^2-a_2^2}\bigg(
a_1^2 \cos\bigg( \frac{\sqrt{c^2 \tau^2-R^2(t')}}{a_1}\bigg)
\nonumber\\
&\quad
-a_2^2 \cos\bigg( \frac{\sqrt{c^2 \tau^2-R^2(t')}}{a_2}\bigg)
-\sqrt{c^2\tau^2-R^2(t')}\,
\bigg(
a_1 \sin\bigg( \frac{\sqrt{c^2 \tau^2-R^2(t')}}{a_1}\bigg)
\nonumber\\
&\hspace*{8cm}
-a_2 \sin\bigg( \frac{\sqrt{c^2 \tau^2-R^2(t')}}{a_2}\bigg)
\bigg)\bigg)
\bigg]
\end{align}
and 
\begin{align}
\label{B-LW-2d}
&B_{(2)}(\bm r,t)
=\frac{\mu_0 qc}{2\pi}\int_{-\infty}^\tR\d t'\, 
\frac{\bm R(t')\times \bm V(t')}{(c^2\tau^2-R^2(t'))^\frac{3}{2}}\, 
\bigg[
1-\frac{1}{a_1^2-a_2^2}\bigg(
a_1^2 \cos\bigg( \frac{\sqrt{c^2 \tau^2-R^2(t')}}{a_1}\bigg)
\nonumber\\
&\quad
-a_2^2 \cos\bigg( \frac{\sqrt{c^2 \tau^2-R^2(t')}}{a_2}\bigg)
-\sqrt{c^2\tau^2-R^2(t')}\,
\bigg(
a_1 \sin\bigg( \frac{\sqrt{c^2 \tau^2-R^2(t')}}{a_1}\bigg)
\nonumber\\
&\hspace*{8cm}
-a_2 \sin\bigg( \frac{\sqrt{c^2 \tau^2-R^2(t')}}{a_2}\bigg)
\bigg)\bigg)
\bigg] \,.
\end{align}
The two-dimensional electromagnetic fields~\eqref{E-LW-2d}
and \eqref{B-LW-2d} draw contributions emitted at all times $t'$ from $-\infty$ up to $\tR$.
The two-dimensional electromagnetic field strengths Eqs.~\eqref{E-LW-2d} and
\eqref{B-LW-2d} are singularity-free.

\subsubsection{1D}

Inserting Eq.~(\ref{J}) into Eqs.~(\ref{E-ret-1d}) and (\ref{B-ret-1d}), 
the spatial integration can be performed to
give the one-dimensional electromagnetic fields in 
generalized Li{\'e}nard-Wiechert form 
\begin{align}
\label{E-LW-1d}
E_{(1)}(x,t)
&=\frac{qc}{2\varepsilon_0(a_1^2-a_2^2)}\int_{-\infty}^\tR\d t' \, 
\frac{ X(t')-\tau V(t') }{\sqrt{c^2\tau^2-X^2(t')}}\,
\bigg[
a_1 J_1 \bigg( \frac{\sqrt{c^2\tau^2-X^2(t')}}{a_1}\bigg)
\nonumber\\
&\hspace*{7cm}
-a_2 J_1 \bigg( \frac{\sqrt{c^2\tau^2-X^2(t')}}{a_2}\bigg)
\bigg]
\,,\\
\label{B-LW-1d}
B_{(1)}(x,t)
&=0\,.
\end{align}
The one-dimensional electric field~\eqref{E-LW-1d}
draws contributions emitted at all times $t'$ from $-\infty$ up to $\tR$.

\section{Self-force, energy release rate and equation of motion of a charged point particle}
\label{sec6}

In this section, we investigate the self-force, the energy release rate and the equation of motion of a charged point particle in 3D in 
the framework of second gradient electrodynamics. 

As in the Maxwell electrodynamics and in the Bopp-Podolsky electrodynamics, 
the Lorentz force is defined by (e.g.~\citep{Post,Jackson,Smith,AM})
\begin{align}
\label{SF}
\bm{\mathcal{F}}=\int_{\Bbb R^3} \big(\rho \bm E +\bm J\times\bm B\big)\, \d V\,.
\end{align}
Inserting Eq.~\eqref{J} into \eqref{SF} and volume integration gives the self-force of a charged particle
\begin{align}
\label{SF-2}
\bm{\mathcal{F}}_{\text{self}}(t)=q \big(\bm E(\bm s(t),t)+\bm V(t)\times\bm B(\bm s(t),t)\big)\,,
\end{align}
which is not zero because of the retardation.

Now substituting the three-dimensional electromagnetic field strengths of a charged point particle~\eqref{E-LW-3d} and \eqref{B-LW-3d}
into Eq.~\eqref{SF-2},
we obtain the explicit expression for the self-force of a charged point particle
\begin{align}
\label{SF-3d}
&\bm{\mathcal{F}}_{\text{self}}(t)
=-\frac{q^2c }{4\pi\varepsilon_0 (a_1^2-a_2^2)}\,
\int_{-\infty}^t\d t'\,
\frac{\bm R(t,t')-\tau \bm V(t')}{(c^2 \tau^2-R^2(t,t'))}\, 
\bigg[
J_2 \bigg(\frac{\sqrt{c^2 \tau^2-R^2(t,t')}}{a_1}\bigg)
\nonumber\\
&\hspace*{9cm}
-J_2 \bigg(\frac{\sqrt{c^2 \tau^2-R^2(t,t')}}{a_2}\bigg)
\bigg]\nonumber\\
&\qquad
+\frac{\mu_0 q^2c}{4\pi(a_1^2-a_2^2)}\,
\bm V(t)\times \int_{-\infty}^{t} 
\d t'\, 
\frac{\bm R(t,t') \times \bm V(t') }{(c^2 \tau^2-R^2(t,t'))}\, 
\bigg[
J_2 \bigg(\frac{\sqrt{c^2 \tau^2-R^2(t,t')}}{a_1}\bigg)
\nonumber\\
&\hspace*{9cm}
-J_2 \bigg(\frac{\sqrt{c^2 \tau^2-R^2(t,t')}}{a_2}\bigg)
\bigg]\,,
\end{align}
where 
\begin{align}
\bm R(t,t')=\bm s(t)-\bm s(t')\,.
\end{align}
In second gradient electrodynamics the self-force~\eqref{SF-3d} is singularity-free.

The (relativistic) Newton equation of motion of a charged particle with charge $q$, position $\bm s(t)$, and velocity $\bm V(t)=\dot{\bm s}(t)$ is
(see, e.g., \citep{CD,Spohn})
\begin{align}
\label{EOM}
\frac{\d}{\d t}\left\{\frac{m_0 \bm V(t)}{\sqrt{1-V^2(t)/c^2}}\right\}=\bm{\mathcal{F}}_{\text{self}}(t)+\bm{\mathcal{F}}_{\text{ext}}(t)\,,
\end{align}
where $m_0$ is the (bare) mass of the particle, 
$\bm{\mathcal{F}}_{\text{self}}(t)$ is the self-force~\eqref{SF-3d} and $\bm{\mathcal{F}}_{\text{ext}}(t)$ is an external force.
If $\bm{\mathcal{F}}_{\text{ext}}(t)$ is given, Eq.~\eqref{EOM}  is an
integro-differential equation for $\bm s(t)$ 
since the self-force~\eqref{SF-3d}  is history dependent. In other words, Eq.~\eqref{EOM} is nonlocal in time due to the 
self-force~\eqref{SF-3d}. 
Only second-order time derivatives of $\bm s(t)$, and no higher derivatives, 
are present in Eq.~\eqref{EOM}. 
In particular, the infamous third-order time-derivative of the position, which exists in the classical Lorentz-Dirac 
equation of motion, does not show up.   
For small particle velocity, Eq.~\eqref{EOM} simplifies to
\begin{align}
\label{EOM-s}
m_0 \, \ddot{\bm s}(t)=\bm{\mathcal{F}}_{\text{self}}(t)+\bm{\mathcal{F}}_{\text{ext}}(t)\,.
\end{align}
Therefore, 
in gradient electrodynamics, the equation of motion of a charged particle
interacting with its own electromagnetic fields is an integro-differential
equation of second order in the time derivative of the particle position 
$\bm s(t)$.

On the other hand, the energy release rate (the rate of change of energy or the electric power) is given by (e.g.~\citep{Post,Jackson,Smith,AM})
\begin{align}
\label{ERR}
{\mathcal{W}}=\int_{\Bbb R^3} \bm J\cdot\bm E \, \d V\,,
\end{align}
which is the rate of doing work or the loss of energy.
Inserting Eq.~\eqref{J} into \eqref{ERR} and volume integration gives the
energy release rate of a charged particle or work done by the self-force  
(electric self-power)
\begin{align}
\label{EER-2}
{\mathcal{W}}_{\text{self}}(t)=q \bm V(t)\cdot\bm E(\bm s(t),t)
=\bm V(t)\cdot \bm {\mathcal{F}}_{\text{self}}(t)
\end{align}
since $\bm V \cdot(\bm V \times\bm B)=0$.
Substituting the three-dimensional electric field strength of a point particle~\eqref{E-LW-3d}
into Eq.~\eqref{EER-2}, we obtain the explicit expression for the 
rate done by the self-force of a charged point particle
\begin{align}
\label{EER-3d}
{\mathcal{W}}_{\text{self}}(t)
&=-\frac{q^2c }{4\pi\varepsilon_0 (a_1^2-a_2^2)}\,
\bm V(t)\cdot
\int_{-\infty}^t\d t'\,
\frac{\bm R(t,t')-\tau \bm V(t')}{(c^2 \tau^2-R^2(t,t'))}\, 
\bigg[
J_2 \bigg(\frac{\sqrt{c^2 \tau^2-R^2(t,t')}}{a_1}\bigg)
\nonumber\\
&\hspace*{8cm}
-J_2 \bigg(\frac{\sqrt{c^2 \tau^2-R^2(t,t')}}{a_2}\bigg)
\bigg]\,. 
\end{align}

The energy equation reads 
\begin{align}
\label{EE}
\frac{\d}{\d t}\left\{\frac{m_0\, c^2}{\sqrt{1-V^2(t)/c^2}}\right\}
={\mathcal{W}}_{\text{self}}(t)+{\mathcal{W}}_{\text{ext}}(t)
\end{align}
with the work due to an external force
\begin{align}
\label{EE-2}
{\mathcal{W}}_{\text{ext}}(t)=\bm V(t)\cdot \bm
{\mathcal{F}}_{\text{ext}}(t)\,. 
\end{align}
Eq.~\eqref{EE} is nonlocal in time due to the 
rate done by the self-force~\eqref{EER-3d}.

\section{Conclusion}
\label{sec7}

\begin{table}[b]
\caption{Behaviour of the Green function of second gradient electrodynamics
  and first-order derivatives on the light cone.}
\begin{center}
\leavevmode
\begin{tabular}{||c|c|c||}\hline
Spatial dimension & Green function $G^{L\square}$ & First-order derivatives of $G^{L\square}$\\
\hline
3D&  approaching zero &   finite and discontinuous\\
2D& approaching zero &  approaching zero\\
1D& approaching zero & approaching zero\\
\hline
\end{tabular}
\end{center}
\label{table}
\end{table}

In this work, we have proposed and developed the second gradient electrodynamics with weak nonlocality in space and time
as an important example of generalized electrodynamics. 
In the framework of second gradient electrodynamics,
the retarded potentials, retarded electromagnetic field strengths, generalized
Li{\'e}nard-Wiechert potentials and electromagnetic field strengths 
in generalized Li{\'e}nard-Wiechert form have been calculated for 3D, 2D and 1D and they depend 
on the entire history from $-\infty$ up to the retarded time $\tR$. 
The electromagnetic field in second gradient electrodynamics is the superposition of the Maxwell field
and a bi-Klein-Gordon field.  
In particular, the bi-Klein-Gordon part gives rise to
an oscillation around the classical Maxwell field. 
The Green function of second gradient electrodynamics $G^{L\square}$ and the first-order
derivatives have been calculated and studied on the light cone (see
Table~\ref{table}) and it has been shown that they are singularity-free 
in 3D, 2D and 1D. 
The Green function $G^{L\square}$ is regular and vanishes on the light cone in 3D, 2D and 1D.
The first-order derivatives of  $G^{L\square}$ vanish on the light cone in 2D and 1D with the exception in 3D which is finite there.  
Therefore, the Green function of second gradient electrodynamics represents a regularization of
the Green function of the d'Alembert equation:
\begin{align}
\label{reg}
G^{L\square}=\text{reg}\, \big[G^\square\big]\,
\end{align}
and its derivatives of first order 
\begin{align}
\label{reg-grad}
\nabla G^{L\square}&=\text{reg}\, \big[\nabla G^\square\big]\,,\\
\label{reg-t}
\pd_t G^{L\square}&=\text{reg}\, \big[\pd_t G^\square\big]\,,
\end{align}
corresponding to the case of the Pauli-Villars regularization 
with two ``auxiliary masses'' $m_1$ and $m_2$.
The  Green function of the bi-Klein-Gordon operator is also singularity-free and it plays the mathematical role of the regularization function 
in second gradient electrodynamics. 
Moreover, the retarded Green functions of second gradient electrodynamics and
their first-order derivatives
show oscillations inside the forward light cone.
In addition, we have calculated the self-force and the equation of motion of a
charged point particle 
and we have shown that the self-force and the energy release rate are singularity-free due to the regular character of the electromagnetic field strengths.
In second gradient electrodynamics,
the equation of motion of a charged point particle is an integro-differential equation 
and the infamous third-order time-derivative of the position does not show up. 
Therefore, second gradient electrodynamics 
represents a singularity-free generalized electrodynamics 
without singularities in the electromagnetic fields on the light cone.

\section*{Acknowledgement}
The author gratefully acknowledges the grant from the 
Deutsche Forschungsgemeinschaft (Grant No. La1974/4-1).

\end{document}